\pdfoutput=1

\documentclass[11pt]{article}

\usepackage[final]{acl}

\usepackage{times}
\usepackage{latexsym}

\usepackage[T1]{fontenc}

\usepackage[utf8]{inputenc}

\usepackage{microtype}

\usepackage{inconsolata}

\usepackage{graphicx}

\usepackage[utf8]{inputenc} 
\usepackage[T1]{fontenc}    
\usepackage{hyperref}       
\usepackage{url}            
\usepackage{booktabs}       
\usepackage{amsfonts}       
\usepackage{nicefrac}       
\usepackage{microtype}      
\usepackage{xcolor}         
\usepackage{xspace}
\usepackage{fdsymbol}
\usepackage{multirow}
\usepackage{multicol}
\usepackage{graphicx}
\usepackage{arydshln}
\usepackage{adjustbox}
\usepackage{wrapfig,lipsum}
\usepackage{subcaption}
\usepackage{color, colortbl}

\usepackage{color, colortbl}
\definecolor{amber}{rgb}{1.0, 0.75, 0.0}
\definecolor{applegreen}{rgb}{0.55, 0.71, 0.0}
\definecolor{treegreen}{rgb}{0.13, 0.54, 0.23}
\definecolor{LightCyan}{rgb}{0.88,1,1}
\newcommand{\bad}{\cellcolor{red!8}}
\newcommand{\worse}{\cellcolor{red!16}}
\newcommand{\good}{\cellcolor{applegreen!10}}
\newcommand{\better}{\cellcolor{applegreen!20}}
\newcommand{\hlcell}{\cellcolor{applegreen!22}}
\newcommand{\gtcell}{\cellcolor{amber!18}}
\newcommand{\hlrow}{\rowcolor{amber!18}}
\newcommand{\dlrow}{\rowcolor{gray!20}}
\newcommand{\dlcell}{\cellcolor{gray!20}}

\usepackage{color-edits}
\addauthor{gn}{magenta}
\addauthor{zw}{orange}
\addauthor{df}{cyan}
\addauthor{vy}{green}
\addauthor{yq}{blue}
\addauthor{fx}{yellow}

\newcommand{\datasetname}{\textsc{CodeRAG-Bench}\xspace}

\title{\datasetname: \\Can Retrieval Augment Code Generation?}

\author{Zora Zhiruo Wang$^{\spadesuit}$\thanks{Equal contribution.} \quad Akari Asai$^{\diamondsuit *}$ \\ {\bf Xinyan Velocity Yu}$^{\heartsuit}$ \quad {\bf Frank F. Xu} $^{\spadesuit}$ \quad {\bf Yiqing Xie} $^{\spadesuit}$ \\ {\bf Graham Neubig} $^{\spadesuit}$ \quad {\bf Daniel Fried} $^{\spadesuit}$ \\
$^{\spadesuit}$Carnegie Mellon University \quad $^{\diamondsuit}$University of Washington \\ $^{\heartsuit}$University of Southern California \\
\url{https://code-rag-bench.github.io/}
}

\begin{document}
\maketitle
\begin{abstract}
While language models (LMs) excel at generating code, many programs are difficult to generate using only parametric knowledge. Despite the success of retrieval-augmented generation (RAG) in text-centric tasks, its potential for code generation remains under-explored. This work introduces \datasetname, a holistic retrieval-augmented code generation benchmark covering tasks like basic programming, open-domain, and repository-level problems and provide reproducible evaluations on both retrieval and end-to-end code generation performance. We further create a diverse, open datastore for code retrieval, aggregating sources such as competition solutions, tutorials, library documentation, StackOverflow posts, and GitHub repositories. Based on \datasetname, we conduct large-scale evaluations of 10 retrievers and 10 LMs and systematically analyze when retrieval can benefit code generation models and identify remaining challenges.  
We find that while retrieving high-quality contexts improves code generation, retrievers often struggle to fetch useful contexts, and generators face limitations in using those contexts effectively. We hope \datasetname~encourages further development in code-oriented RAG methods.
\end{abstract}

\section{Introduction}
\label{sec:1:intro}

Generating code from natural language has rapidly advanced with language models (LMs; \citealt{chen2021evaluating,yujia2022competition,li2023starcoder,roziere2023code}). However, most models follow an NL (Natural Language)-to-code approach without integrating external context, which is crucial in complex scenarios like using unfamiliar libraries \citep{zhou2023docprompting,jimenez2024swebench}. Relying solely on parametric knowledge also limits adaptation to new data distributions at test time, such as evolving public libraries or private code bases not seen during training~\cite{zhang2023repocoder,jimenez2024swebench}.

Retrieval-augmented generation (RAG; \citealt{lewis2020retrieval,guu2020retrieval}) addresses this by retrieving relevant documents at inference time, reducing reliance on model parameters~\cite{asai2024reliable} and improving accuracy across tasks \cite{Izacard2022FewshotLW}. Despite success in text-based tasks, its application to diverse coding problems and retrieval sources remains under-explored~\citep{zhou2023docprompting,su2024arks}.

\begin{figure*}[t!]
\vspace{-5mm}
    \centering
    \includegraphics[width=0.92\textwidth]{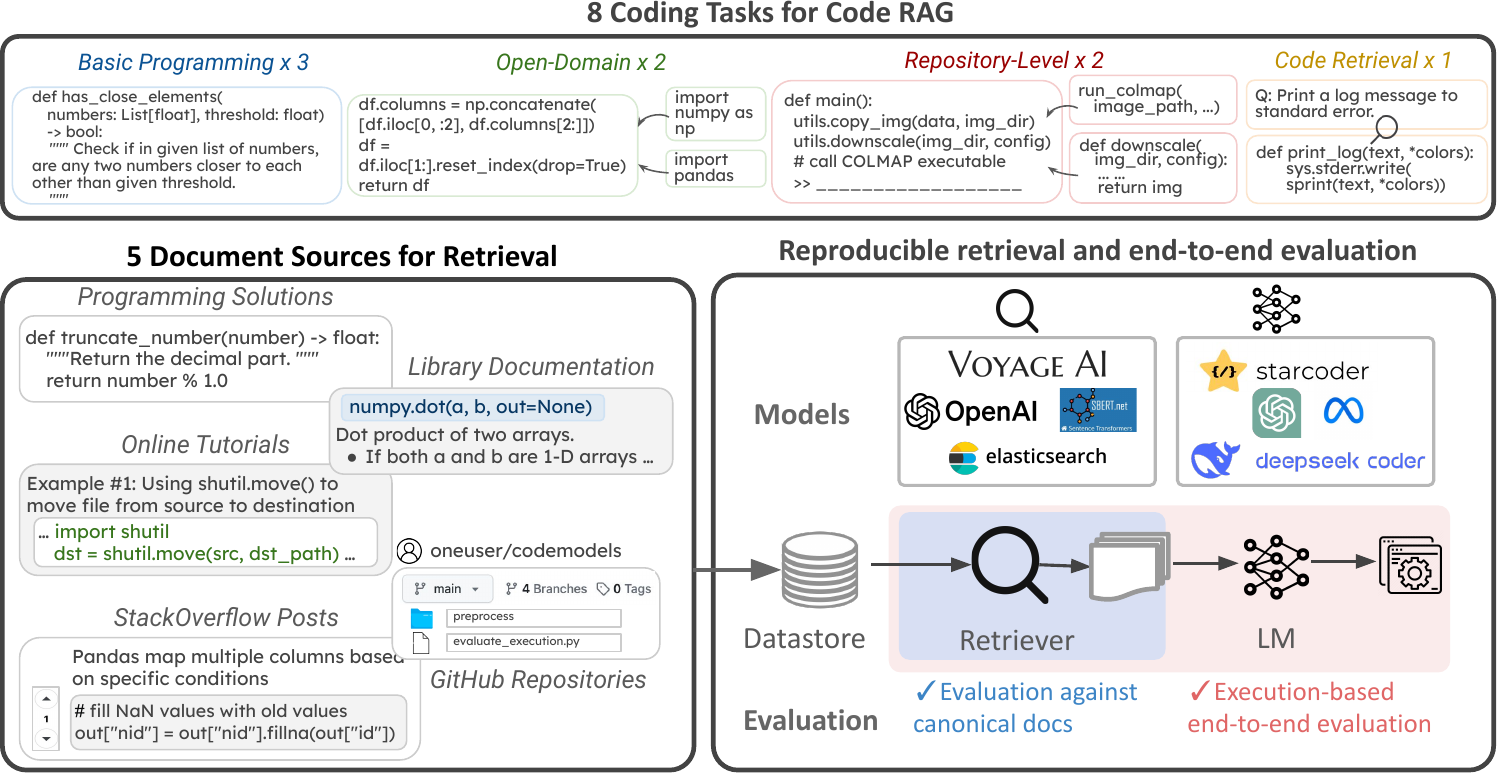}
    \vspace{-2mm}
    \caption{Overview of \datasetname.}
    \label{fig:overview}
\vspace{-4mm}
\end{figure*}

We present \datasetname, a holistic benchmark designed to advance research in retrieval-augmented code generation (RACG; \S\ref{sec:2:benchmark}). 
\datasetname~(as in Figure~\ref{fig:overview}) covers six programming tasks across four categories: basic programming, open-domain coding, repository-level, and code retrieval tasks. For each task, we manually annotate canonical documents as references for evaluating RACG systems. We also compile a diverse corpus of documents from five sources: programming solutions, online tutorials, Python library documentation, StackOverflow posts, and GitHub files. In total, \datasetname has 9k coding tasks and 25 million retrieval documents, providing a robust foundation for reproducible and reliable evaluations in retrieval and RACG. 

We conduct holistic evaluations in retrieval, generation, and RACG (\S\ref{sec:3:baseline}). 
Code generation models significantly benefit from access to canonical documents (i.e., from the canonical retrieval corpus) in various scenarios. For example, GPT-4o achieves a 27.4\% gain on SWE-Bench and a 6.9\% gain on the harder ODEX subset when canonical documents are provided. In RACG settings, where models retrieve top relevant documents, some even surpass their performance when using gold documents, highlighting the strong potential of retrieval-augmented approaches for enhancing code generation. However, current retrieval models face challenges in selecting useful documents, particularly for open-domain and repository-level tasks. Additionally, generation models with limited context windows exhibit smaller improvements, suggesting considerable room for future advancements.

Beyond canonical retrieval, we also explore RACG with open retrieval, i.e., retrieving documents from various sources with different chunking strategies (\S\ref{sec:4:retrieve-sources}). 
We find that models can benefit from functionally relevant snippets from certain sources, and chunking documents to 200--800 tokens often gives the best results. For instance, by retrieving from StackOverflow or online tutorials, both StarCoder and GPT4o can significantly improve, while on repository-level tasks, the gains are rather limited.  
Overall, we hope \datasetname~can serve as a testbed for future work exploring, analyzing, and improving RACG systems. 
\section{The \datasetname}
\label{sec:2:benchmark}

For \datasetname (Figure~\ref{fig:overview}), the curation is driven by three factors: (i) \textbf{Diverse tasks}: Code generation spans multiple levels (line, function, repository) across closed and open domains. (ii) \textbf{Rigorous evaluation}: We offer high-quality ground-truth annotations for retrieval and execution-based evaluation to measure functional correctness. (iii) \textbf{Unified interface}: Our codebase provides a consistent interface for retrieval, augmented generation, and evaluation, unlike current datasets with varied pipelines. 

In this section, we introduce the creation process of \datasetname: programming problem integration (\S\ref{sec:2.1:programming-problems}), retrieval source collection (\S\ref{sec:2.2:retrieval-sources}), canonical document annotation (\S\ref{sec:2.3:canonical-document}), and the evaluation pipeline (\S\ref{sec:evaluation_metrics}). 
Examples with canonical documents are available in \S\ref{app:experiment-details}.

\begin{table*}[t!]
\small
\centering
\vspace{-4mm}
\resizebox{0.92\textwidth}{!}{
  \begin{tabular}{ll|cccc}
    \toprule
    \multicolumn{1}{c}{\textbf{Type}} & \multicolumn{1}{c|}{\textbf{Dataset}} & \textbf{\# Examples} & \textbf{\# Corpus} & \textbf{Ground-Truth Docs} & \textbf{Evaluation} \\
    \midrule
    \multirow{3}{*}{Basic programming} & {HumanEval} & {~~164} & {~~164} & {program solutions} & {execution}   \\
    {} & {MBPP} & {~~500} & {~~500} & {program solutions} & {execution} \\
    {} & {LiveCodeBench} & {~~400} & {-} & {-} & {execution} \\
    \midrule
    \multirow{2}{*}{Open-domain} & {DS-1000} & {1000} & {34,003} & {docs} & {execution}   \\
    {} & {ODEX} & {~~945} & {34,003} & {docs, stackoverflow} & {execution} \\
    \midrule
    \multirow{2}{*}{Repository-level} & {RepoEval (function)} & {~~373} & {~~237} & {github repository} & {execution}   \\
    {} & {SWE-bench-Lite} & {~~300} & {40,868} & {github repository} & {execution} \\
    \midrule
    {Code retrieval} & {CodeSearchNet-Py} & 22,177 & {22177} & {CSN functions} & {ndcg@10} \\
    \bottomrule
  \end{tabular}
}
\vspace{-1mm}
\caption{Overview of the datasets in CodeRAG-Bench. CSN stands for CodeSearchNet.}
\label{tab:benchmark-overview}
\vspace{-3mm}
\end{table*}

\subsection{Programming Problems}
\label{sec:2.1:programming-problems}
We categorize existing Python-based coding datasets into four types:\footnote{In this work we focus on Python-related tasks because it is the most widely-used programming language for benchmarking code generation. We leave extensions to other programming languages for future work.} code retrieval, basic programming, open-domain problems, and repository-level problems. 
To ensure the diversity of datasets, we choose and unify multiple frequently adopted datasets for each category, as listed in \autoref{tab:benchmark-overview}.

\noindent \textbf{Basic programming problems} \quad
This category includes interview-style problems that mostly require Python built-in operations and pose algorithmic challenges. 
We select the two most widely used datasets: HumanEval \citep{chen2021evaluating} and MBPP \citep{austin2021program}, which ask the model to complete a function from an NL problem description. 
However, due to limited public knowledge about model training data, it is unclear whether models suffer from data contamination on HumanEval and MBPP~\citep{jain2024livecodebench}. Hence, we also include LiveCodeBench \citep{jain2024livecodebench} with problems collected from coding websites after the training cutoff of LMs that we consider, to decrease the risk of contamination.

\noindent \textbf{Open-domain problems} \quad
Open-domain coding problems require Python libraries beyond the standard libraries used in basic programming problems.
We adopt the DS-1000 \citep{lai2023ds} and ODEX \citep{wang2023execution} datasets that cover data-science and general open-domain coding problems. 
DS-1000 collects data science problems with programs using seven common data-related libraries such as \texttt{pandas} and \texttt{numpy}.
ODEX covers problems using a broader range of 79 libraries, such as web requests with \texttt{requests} and database operations with \texttt{sqlalchemy}.

\noindent \textbf{Repository-level coding problems} \quad
Beyond function-level, some problems require editing files in the context of an entire GitHub repository.
We thus adopt RepoEval \citep{zhang2023repocoder} and SWE-bench~\citep{jimenez2024swebench} for repository-level code generation and issue-solving tasks. 
We integrate all three splits of RepoEval but only report its function split, as it is the only split supporting execution-based evaluation.\footnote{Two other splits (API and line) are evaluated by lexical measures that have been shown as ineffective in signifying functional correctness \citep{chen2021evaluating,wang2023execution}.}
Notably, our codebase is the first to enable reproducible execution evaluation on RepoEval.
SWE-bench focuses on resolving GitHub issues by asking models to edit multiple files that pass the required test cases. 
We use SWE-bench-Lite,\footnote{\url{https://www.swebench.com/lite.html}} a 300-problem subset whose results can be reproduced, with a packaged Docker container \citep{wang2024opendevin}.

\noindent \textbf{Code retrieval problems} \quad
In addition to retrieval for augmenting generations, we adopt the Python split of CodeSearchNet (CSN) as a code retrieval task. CSN searches for the correct implementation of an NL query from a pool of functions collected from GitHub repositories. 
Instead of monitoring how generation changes with various retrieval results, CSN can directly measure retrieval quality.


\subsection{Retrieval Sources}
\label{sec:2.2:retrieval-sources}
We collect retrieval documents from five commonly used resources for program developers, listed in \autoref{tab:retrieval-sources}. 
{\datasetname~ supports two retrieval setups: {\bf canonical retrieval}---retrieves documents from only the canonical datastore (\S\ref{sec:2.3:canonical-document}), and {\bf open retrieval}---retrieves documents from any datastore.}

\noindent \textbf{Programming solutions} \quad
We create one document from each basic programming problems that have canonical solutions (i.e., HumanEval and MBPP), following \citet{voyage2024voyage}, by concatenating its NL problem and program solution.

\noindent \textbf{Online tutorials} \quad
We collect tutorials from multiple websites including GeeksforGeeks, W3Schools, tutorialspoint, and Towards Data Science,\footnote{\url{https://geeksforgeeks.org}; \url{https://www.w3schools.com/}; \url{https://www.tutorialspoint.com/}; \url{https://towardsdatascience.com}} via the raw HTML pages obtained from ClueWeb22~\cite{overwijk2022clueweb22}, a large-scale crawled web corpus.
Each page contains code snippets and their text explanations, covering topics from basic programming techniques to advanced library usage. 

\noindent \textbf{Library documentation} \quad
We collect the official documentation provided by \url{devdocs.io} for all Python libraries following \citep{zhou2023docprompting}. These could be especially useful for open-domain and repository-level problems that use some library functions to realize complex setups.

\noindent \textbf{StackOverflow posts} \quad
StackOverflow (SO) is among the most frequently visited sites for developers.
We collect all SO posts from the RedPajama-1T~\cite{together2023redpajama} \texttt{stackexchange} split. We treat each post as a retrievable document, that has a question, code responses, and textual explanations. 

\noindent \textbf{GitHub repository} \quad
We collect high-quality repositories from GitHub, using the \texttt{github} split of RedPajama-1T~\cite{together2023redpajama}, as developers often refer to popular repositories when writing their programs. 
Following this practical paradigm, we enable LMs to retrieve files from other repositories as contexts to write the current program.

\begin{table}[ht]
\small
\centering
\vspace{-1mm}
\begin{tabular}{lrr}\\\toprule  
\multicolumn{1}{c}{\textbf{Resource}} & \multicolumn{1}{c}{\textbf{Corpus size}} & \multicolumn{1}{c}{\textbf{Avg. length}} \\
\midrule
Programming solutions & 1.1$k$ & 194.6 \\
Online tutorials & 79.4$k$ & 1502.5 \\
Library documentation & 34$k$ & 953.4 \\
StackOverflow posts & 23.5$M$ & 689.2 \\
Github files & 1.7$M$ & 5135.4 \\
\bottomrule
\end{tabular}
\vspace{-2mm}
\caption{Five sources to form our retrieval datastore.}
\label{tab:retrieval-sources}
\vspace{-3mm}
\end{table}

\subsection{Canonical Document Annotation}
\label{sec:2.3:canonical-document}
To ensure reliable retrieval evaluation and estimate the upper bound of a RACG system with an ideal retriever, it's essential that all examples include \emph{canonical documents}--the documents containing the necessary context to solve the programming problem. As most existing datasets lack these canonical documents, we annotate them from the corresponding retrieval pool, as shown in \autoref{tab:benchmark-overview}. 

\noindent \textbf{Basic programming problems}  \quad
The canonical document for examples in HumanEval and MBPP is the documents we created in \S\ref{sec:2.2:retrieval-sources} in the \textit{programming solutions} pool.
Since LiveCodeBench does not provide solutions to its problems, we do not annotate canonical documents for it.

\noindent \textbf{Open-domain problems} \quad
Since open-domain problems require libraries, we annotate the canonical \textit{library documentation} for DS-1000 and ODEX examples. 
We first automatically parse out the library functions used in each program, and find their corresponding documentation entries. Then, we manually verify the functions and remove incorrect ones. This yields an average of 1.4 and 1.2 entries for DS-1000 and ODEX.

\noindent \textbf{Repository-level problems} \quad
We adopt \emph{canonical code} from the original dataset as our canonical documents: 20-line code snippets of the missing functions in RepoEval, and the ground-truth edited files in SWE-bench. We obtain these from the completed local repositories from the original datasets.

\subsection{Evaluation Metrics}
\label{sec:evaluation_metrics}
\vspace{-1mm}
For retrieval, we evaluate NDCG, Precision and Recall~\cite{thakur2021beir} and use  NDCG@10 percentage as our primary metric, following prior work~\cite{izacard2022unsupervised}.  
For code generation, we adopt the pass@k metric \citep{chen2021evaluating} to measure the execution correctness of programs. {We evaluate the final RAG performance both in canonical and open retrieval setups.}

\begin{table*}[t!]
\vspace{-4mm}
\small
\begin{center}
\resizebox{0.98\textwidth}{!}{
  \begin{tabular}{l|ccc|cc|cc|c}
    \toprule
    \multicolumn{1}{c|}{\multirow{2}{*}{\textbf{Method}}} & \multicolumn{3}{c|}{\textbf{Problem Solutions}} & \multicolumn{2}{c|}{\textbf{Library Docs}} & \multicolumn{2}{c|}{\textbf{In-Repository Files}} & {\bf Avg.}  \\
    {} & {HumanEval} & {MBPP} & {CSN} & {DS-1000} & {ODEX} & {RepoEval} & {SWE-bench-Lite} & All \\
    \midrule
    {BM25} & {\bf 100.0} & {98.6} & {89.1} & {5.2} & {6.7} & {93.2} & {43.0} & 57.7 \\
    \midrule
    {GIST-base ($768$)} & {98.0} & {98.0} & {89.9} & {12.0} & {12.1} & {81.2} & {46.8} & 58.0  \\
    {GIST-large ($1024$)} & {\bf 100.0} & \underline{98.9} & {89.6} & {13.6} & \underline{28.0} & {82.9} & {47.8} & 61.7 \\
    {BGE-base ($768$)} & \underline{99.7} & {98.0} & {90.0} & {10.8} & {22.0} & {77.5} & {44.9}  & 58.8 \\
    {BGE-large ($1024$)} & {98.0} & {\bf 99.0} & \underline{90.6} & {8.9} & {11.5} & {80.4} & {40.1} & 56.3  \\
    {SFR-Mistral ($4096$)} &  \bf 100.0 & \bf 99.0 & -&  19.3 & \bf 37.1 & 83.8 & {\bf 62.7} & {\bf 67.0}  \\
    \rowcolor{LightCyan}
    {Codesage-small ($768$)} & {100.0} & {96.3} & {\bf 90.7} & {8.9} & {14.3} & \underline{94.1} & {47.1} & 60.1    \\
    \rowcolor{LightCyan}
    {Jina-v2-code ($768$)} & {\bf 100.0} & {97.7} & {-} & \underline{26.2} & {19.9}& {90.5} & \underline{58.3} & \underline{65.4}   \\
    \midrule
    {OpenAI-03 ($1536$)} & {\bf 100.0} & \underline{98.9} & {-} & {18.2} & {16.5}& {93.0} & {43.3} & 61.7  \\
    \rowcolor{LightCyan}
    {Voyage-code ($1536$)} & {\bf 100.0} & {\bf 99.0} & {-} & {\bf 33.1} & {26.6} & {\bf 94.3} & {29.1}  & 63.7  \\
    \bottomrule
  \end{tabular}
}
\end{center}
\vspace{-3mm}
\caption{Retrieval performance (NDCG@10) on code generation datasets. LiveCodeBench is excluded due to lack of ground-truth solutions. RepoEval is at the function level with 2k context tokens. Embedding dimension sizes are listed next to method names. Bold indicates best performance, underline indicates second-best. Highlighted models are specifically trained for code domains. Avg. reflects overall scores, excluding CodeSearchNet. }
\label{tab:baseline-retrieval-results}
\vspace{-2mm}
\end{table*}

\section{Canonical RACG} 
\label{sec:3:baseline}
\vspace{-2mm}

We evaluate 10 top retrieval and 10 generation models on \datasetname with canonical data sources. 
We report results of document retrieval (\S\ref{sec:3.2:retrieval-results}), direct NL-to-code generation (\S\ref{sec:3.3:generation-results}), and end-to-end RACG with retrieved context (\S\ref{sec:3.4:crag-results}).

\subsection{Experimental Setup}
\label{sec:3.1:experiment-setup}
\vspace{-1mm}

\noindent \textbf{Retrieval baselines} \quad
We adopt 10 top-performing retrievers from three categories: sparse, dense, and proprietary APIs. For sparse retrievers, we use BM25~\cite{Robertson2009ThePR}, known for its robustness in domain adaptation~\cite{thakur2021beir}. Dense retrievers include BGE-base/large~\citep{bge_embedding}, GIST-base/large~\citep{solatorio2024gistembed}, and SFR-Embedding-Mistral~\citep{SFRAIResearch2024}, all top-ranked on the MTEB leaderboard~\citep{muennighoff2022mteb}. We also include open code embedding models, Codesage-small~\cite{zhang2024code} and Jina-v2-code~\citealp{gunther2023jina}, which are specifically trained for code retrieval. 
Proprietary APIs include \texttt{voyage-code-2}~\citep{voyage2024voyage}, optimized for code retrieval, and \texttt{openai-text-embedding-small-03}, selected for its cost-effectiveness. Finally, we apply reranking with BGE-reranker-base\citep{bge_embedding} on top-100 \texttt{openai} results before generation. 

\noindent \textbf{Generation baselines} \quad
We adopt both code-specific LMs and strong general text-oriented LMs.
For code-specific LMs, we use StarCoder2 \citep{lozhkov2024starcoder}, CodeGemma \citep{codegemma2024}, CodeLlama \citep{roziere2023code}, and DeepSeekCoder \citep{guo2024deepseek} in various sizes.
For general text LMs, we include three top-performing models: Llama3 \citep{introducing2024}, Command-R \citep{command-r} specially optimized for RAG, and proprietary GPT models \texttt{gpt-3.5-turbo-0125} and \texttt{gpt-4o}.
We use the instruct version of all generation models if available, since they often perform better than the base versions.

\noindent \textbf{Experimental setup} \quad
For retrieval, we implement BM25 retrievers using pyserini \citep{Lin_etal_SIGIR2021_Pyserini} with parameter $k_1 = 1.2$ and $b = 0.75$, and use {sentence-transformers}~\cite{reimers-2019-sentence-bert}\footnote{\url{https://sbert.net/}} for all dense models with open checkpoints. 
We prepend the top-5 retrieved documents to the original problems (we study the number of documents in \S\ref{app:num-docs}), and do not include other contexts such as few-shot examples. 
For code generation, we use temperature $t=0.2$, $top\_p=0.95$ and sample one response for all generations, following prior work~\citep{li2023starcoder}. 
Specifically on SWE-bench-Lite, we adopt the $n=21$ way sampling and majority-vote reranking strategy proposed by Agentless \citep{xia2024agentless}.\footnote{We found that without these approaches, performance even with state-of-the-art GPT4o remains around 1-2\%.}


\subsection{Retrieval Results}
\label{sec:3.2:retrieval-results}

\autoref{tab:baseline-retrieval-results} shows retrieval results on six tasks.

\noindent \textbf{Comparison of lexical and neural retrievers} \quad
BM25 has been widely used as a primary retrieval model in recent RACG work~\citep{zhou2023docprompting,jimenez2024swebench}, yet comprehensive comparisons against diverse retrieval systems are often under-explored. 
While prior studies indicate that neural retrieval systems often underperform BM25 baselines in out-of-domain scenarios~\citep{thakur2021beir}, 
our analysis of \datasetname reveals that dense embedding models frequently surpass BM25. 
We hypothesize that this is because many competitive retrieval models are trained on diverse tasks across various domains, including code data ~\citep{asai-etal-2023-task,su-etal-2023-one}, enhancing their robustness in code retrieval setups.

\noindent \textbf{Do code retrieval models perform better?}\quad
At similar parameter scales, models specifically trained for code retrieval tasks typically show superior performance. Notably, Jina-v2-code outperforms GIST-base and BGE-base by 7.4 and 6.6 average NDCG@10, respectively, while Voyage-code significantly outperforms OpenAI-03.

\noindent \textbf{Do larger retrieval models perform better?} \quad
Among dense retrieval models, increasing model size often leads to better retrieval performance, similar to the trends observed in LMs~\cite{Brown2020LanguageMA}. 
In particular, GIST-large (340$M$) constantly outperforms GIST-base (110$M$), and SFR-Mistral (7$B$) 
achieves the best among all open sparse and dense models on all tasks, surpassing proprietary embedding models on several tasks. 

\begin{table}[ht]
\vspace{-4mm}
\small
\resizebox{\linewidth}{!}{
\begin{tabular}{lrrrr}\\\toprule  
Method & \multicolumn{1}{c}{\textbf{Encoding}} & \multicolumn{1}{c}{\textbf{Search}} & \multicolumn{1}{c}{\textbf{Model}} & 
\multicolumn{1}{c}{\textbf{Index}} \\
\midrule
BM25 & 0.15ms & 0.02ms & - & 141MB \\
GIST-base & 3.7ms & 9.7ms & 440MB & 307MB \\ 
GIST-large & 13ms & 18ms & 1300MB & 409MB \\
SFR-Mistral & 316ms & 113ms & 14220MB &1638 MB\\
\dlrow {Voyage-code} & 22ms & 40ms & - & 1172MB \\ 
\dlrow {OpenAI-03} & 31ms & 47ms & - & 1172MB \\
\bottomrule
\end{tabular}
}
\vspace{-1mm}
\caption{Efficiency analysis for document retrieval.}
\label{tab:efficiency}
\vspace{-1mm}
\end{table}

\noindent \textbf{Efficiency} \quad
While larger retrieval models often outperform smaller ones, they often introduce significant costs. We analyze efficiency, focusing on (i) {\it encoding latency}: latency to encode documents offline, and (ii) {\it search latency}: latency to encode queries/documents and calculate their similarities, (iii) {\it model storage requirements}, and (iv) {\it index storage requirements}. 
We conduct efficiency analysis on sampled CodeSearchNet Python data.\footnote{Due to the costs, we randomly sample 10k queries and 100k  from CodeSearchNet Python split. For API models, we use a batch size of 64 for encoding.} See experimental details in \S\ref{app:retrieval-cost}. As shown in \autoref{tab:efficiency}, BM25 indexing and searching takes only seconds to finish.
Compared to base-size GIST-base, the SFR-Mistral model is more powerful in retrieval, yet requires over 5$\times$ larger index storage, and adds nearly 100$\times$ latency to encode documents, suggesting that the efficiency aspect should also be carefully studied for RAG pipelines.

\begin{table*}[ht]
\vspace{-5mm}
\small
\centering
\resizebox{\textwidth}{!}{
  \begin{tabular}{l|cc|cc|c|cc|cc|cc|cc|cc}
    \toprule
    \multicolumn{1}{c|}{\multirow{3}{*}{\textbf{Method}}} & \multicolumn{5}{c|}{\textbf{Basic Programming}} & \multicolumn{6}{c|}{\textbf{Open-Domain}} & \multicolumn{4}{c}{\textbf{Repo-Level}} \\
    {} & \multicolumn{2}{c}{HumanEval} & \multicolumn{2}{c}{MBPP} & {LCB} & \multicolumn{2}{c}{DS-1000} & \multicolumn{2}{c}{ODEX} & \multicolumn{2}{c|}{ODEX-hard} & \multicolumn{2}{c}{RepoEval} & \multicolumn{2}{c}{SWE-bench} \\
    {} & {w/o} & {gold} & {w/o} & {gold} & {w/o} & {w/o} & {gold} & {w/o} & {gold} & {w/o} & {gold} & {w/o} & {gold} & {w/o} & {gold} \\
    \midrule
    {StarCoder2-7B} & 31.7 & \better{94.5} & 10.4 & \better{34.8} & {~~1.5} & 29.2 & \good{30.0} & 14.6 & \good{17.5} & 10.3 & \good{17.2} & 26.5 & \better{42.0} & 0.0 & {~~0.7} \\
    {CodeGemma-7B} & 49.4 & \better{77.4} & 48.0 & \good{52.2} & {21.5} & 20.1 & \bad{19.8} & 18.9 & \bad{18.2} & 13.8 & {13.8} & 24.7 & \good{32.2} & 0.0 & {~~0.3} \\
    {CodeLlama-7B} & 34.8 & \better{87.2} & 23.8 & \better{42.8} & {13.5} & 21.8 & \good{26.1} & 35.8 & \good{41.0} & 27.6 & \good{31.0} & 24.1 &  \better{38.3} & 0.0 & {~~0.0} \\
    {CodeLlama-34B} & 42.7 & \better{84.8} & 51.2 & \better{88.0} & {~~5.8} & 34.7 & \good{37.0} & 34.9 & \good{38.0} & 17.2 & \better{27.6} & 29.8 & \better{42.6} & 0.0 & {~~0.0} \\
    {DeepSeekCoder-7B} & 70.1 & \better{87.8} & 60.8 & \good{63.6} & {30.5} & 41.4 & \good{43.2} & 39.2 & \good{41.7} & 17.2 & \good{24.1} & 28.2 &  \better{43.7} & 0.0 & {~~0.0} \\
    {DeepSeekCoder-33B} & 78.0 & \better{95.7} & 61.0 & \better{92.2} & {33.8} & 40.2 & \bad{40.1} & 28.0 & \good{28.9} & 24.1 & \good{31.0} & 32.4 & \better{45.3} & 0.3 & {~~0.7} \\
    \midrule
    {Llama3-8B} & 57.9 & \good{65.2} & 35.6 & \better{52.8} & {~~2.8} & 28.9 & \good{31.1} & 37.4 & \bad{33.7}  & 13.8 & \good{17.2} & 26.0 & \better{43.2}  & 0.0 & {~~0.3} \\
    {Command-R} & 43.3 & \good{51.2} & 37.2 & \good{37.8} & {10.0} & 25.8 &  \good{28.5} & 35.5 & \good{36.0} & 20.7 & {20.7} & 23.9 & \better{37.0} & 0.0 & {~~0.3} \\
    {GPT-3.5-turbo} & 72.6 & \better{91.5} & 70.8 & \good{72.6} & {35.3} & 43.7 & \bad{42.9} & 41.7 & \bad{40.3} & 17.2 & \good{24.1} & 23.9 & \better{39.1} & 0.7 & \good{~~6.3} \\
    {GPT-4o} & 75.6 & \better{92.6} & 79.4 & \good{81.4} & {43.8} & 52.7 &  \bad{51.2} & 44.6 & \bad{44.2} & 20.7 & \good{27.6} & 32.4 & \better{46.1} & 2.3 & \better{30.7} \\
    \bottomrule
  \end{tabular}
}
\vspace{-2mm}
\caption{Code generation pass@1 without additional contexts (\textit{w/o}), and with ground-truth documents (\textit{gold}).  We only report \textit{w/o} for LCB  because LCB does not have ground-truth documents. We highlight results showing \textit{gold} > \textit{w/o} with green (darker green when having 10+ increases), and with red if \textit{gold} < \textit{w/o}.}
\label{tab:baseline-generation-results}
\vspace{-3mm}
\end{table*}

\subsection{Generation with Canonical Documents}
\label{sec:3.3:generation-results}
We first evaluate possible lower- and upper-bounds on RACG results by testing generation (i) without any retrieval, and (ii) with ground-truth documents. We report both results in \autoref{tab:baseline-generation-results}.
Compared to the base generation without contexts, incorporating canonical contexts improves in most setups, and substantially so on \textit{basic programming} problems.

\noindent \textbf{On open-domain tasks}, most code-specific LMs increase up to $5.2$ points, signifying that most models can benefit from indirectly helpful documents. 
In contrast, GPTs show no gains with retrieval. We hypothesize that this is because both datasets mostly test on common Python libraries, which powerful models may have already memorized, similar to their memorization of factual knowledge \citep{mallen-etal-2023-trust,kandpal2023large}, thereby reducing the need for retrieval. 
To verify this hypothesis, we build an ODEX subset of examples with the 20 least used libraries, i.e., {\bf ODEX-hard}. 
As shown in \autoref{tab:baseline-generation-results}, adding documents retrieved with most methods improves the results by 20.3--40.1\%, showing the effectiveness of RACG on challenging coding tasks using unfamiliar libraries.

\noindent \textbf{Repository-level challenges} All models show gains of 7.5–17.2 points with canonical snippets in RepoEval, but SWE-bench Lite proves much more challenging --- only GPT-3.5-turbo and GPT-4o achieve non-trivial results, consistent with previous findings \citep{yang2024swe}. Notably, GPT-4o shows a 27.4\% increase when using gold documents on SWE-bench, indicating that retrieval significantly enhances performance when paired with strong core generation capabilities, even in highly challenging coding tasks.

\subsection{Retrieval-Augmented Code Generation}
\label{sec:3.4:crag-results}

We now experiment with top-performing retrieval and generation models in the full RACG setting, which requires both retrieve documents and generating conditioned on the documents.
We select the best retrieval models from each type: BM25, GIST-large, Voyage, and OpenAI embeddings.
For generation, we select (i) StarCoder2-7B: a weaker model that benefits the most from contexts; (ii) DeepSeekCoder-7B: one of the strongest open code LMs; and (iii) GPT-3.5-turbo: one of the top proprietary models. 
 For each dataset, we retrieve the most relevant contexts from its canonical source marked in \autoref{tab:benchmark-overview}, 
 and retrieve \textit{programming solutions} for LiveCodeBench.
\autoref{tab:baseline-rag-results} shows the results. 
Note that we exclude canonical docs (answers) from the retrieval corpora for basic programming tasks. 

Overall, the best retrieval models vary depending on the task and underlying LMs. In some cases, top-performing retrieval models do not lead to the best RACG outcomes, highlighting the need to evaluate RACG systems holistically across varied tasks. 

\begin{table*}[ht]
\vspace{-4mm}
\small
\centering
\resizebox{0.88\textwidth}{!}{
  \begin{tabular}{l|ccc|ccc|cc}
    \toprule
    \multicolumn{1}{c|}{\multirow{2}{*}{\textbf{Method}}} & \multicolumn{3}{c|}{\textbf{Basic Programming}} & \multicolumn{3}{c|}{\textbf{Open-Domain}} & \multicolumn{2}{c}{\textbf{Repo-Level}} \\
    {} & {HumanEval} & {MBPP} & {LCB} & {DS-1000} & {ODEX} & {ODEX-hard} & {RepoEval} & {SWE-bench} \\
    \midrule
    \multicolumn{9}{c}{\textit{w/ StarCoder2-7B}} \\
    \midrule
    \dlrow {None} & {31.7} & {2.4} & {1.5} & {29.2} & {14.6} & {10.3} & {26.5} & {0.0} \\
    {BM25} & \better{\bf 43.9} & \better{51.8} & {1.0} & \good{\bf 36.7} & \bad{14.1} & \good{13.8} & \better{36.7} & {0.0} \\
    {GIST-large} & \good{38.7} & \better{50.4} & {0.5} & \good{35.9} & \good{\bf 17.3} & \good{13.8} & \better{40.8} & {0.3} \\
    {Voyage, code} & \good{39.0} & \better{52.6} & {0.3} & \good{36.0} & \good{15.3} & {10.3} & \better{45.8} & {0.3} \\
    {OpenAI, small} & \good{39.0} & \better{52.6} & {\bf 1.5} & \good{35.5} & \good{15.9} & \good{\bf 17.2} & \better{51.2} & {0.0} \\
    {\it OpenAI, rerank} & \good{34.8} & \better{\bf 53.4} & {0.5} & \good{33.4} & \bad{14.1} & \good{\bf 17.2} & \better{\bf 53.9} & {0.3} \\
    \hlrow {Gold} & {94.5} & {34.8} & {-} & {30.0} & {17.5} & {17.2} & {42.0} & {0.7} \\
    \midrule
    \multicolumn{9}{c}{\textit{w/ DeepseekCoder-7B-instruct}} \\
    \midrule
    \dlrow {None} & {70.1} & {60.8} & {30.5} & {41.4} & {39.2} & {17.2} & {28.2} & {0.7} \\
    {BM25} & \bad{\bf 68.9} & {60.0} & \good{31.8} & \bad{\bf 36.6} & \bad{37.8} & \good{20.7} & \good{37.3} & {0.0} \\
    {GIST-large} & \bad{66.3} & \bad{56.6} & \good{\bf 33.8} & \bad{35.9} & \bad{34.9} & \good{20.7} & \better{44.5} & {0.3} \\
    {Voyage, code} & \bad{66.5} & \bad{56.4} & \good{31.8} & \bad{35.9} & \good{\bf 39.4} & {17.2} & \better{46.6} & {0.3} \\
    {OpenAI, small} & \bad{\bf 68.9} & \bad{58.6} & \good{32.0} & \bad{35.5} & \bad{37.1} & \good{20.7} & \better{55.2} & {0.3} \\
    {\it OpenAI, rerank} & \bad{53.0} & {\bf 60.6} & \good{31.5} & \bad{36.5} & \bad{37.1} & \good{\bf 24.1} & \better{\bf 55.5} & {0.3} \\
    \hlrow {Gold} & {87.8} & {63.6} & {-} & {43.2} & {41.7} & {24.1} & {48.1} & {0.0} \\
    \midrule
    \multicolumn{8}{c}{\textit{w/ GPT-3.5-turbo}} & \textit{GPT-4o} \\
    \midrule
    \dlrow {None} & {72.6} & {70.8} & {35.3} & {43.7} & {41.7} & {17.2} & {23.9} & {~~2.3} \\
    {BM25} & \good{73.2} & \good{72.4} & \good{\bf 35.5} & \bad{36.9} & \bad{\bf 41.0} & \good{\bf 24.1} & \good{30.8} & \good{~~6.7} \\
    {GIST-large} & \good{73.2} & \bad{68.2} & \bad{34.8} & \bad{36.7} & \bad{36.2} & \bad{13.8} & \better{38.3} & \better{19.3} \\
    {Voyage, code} & \good{\bf 75.0} & \bad{66.8} & \bad{34.5} & \bad{\bf 37.4} & \bad{\bf 41.0} & \good{20.7} & \better{43.2} & \better{15.7} \\
    {OpenAI, small} & \good{73.8} & \bad{68.4} & \good{35.8} & \bad{36.9} & \bad{40.3} & {17.2} & \better{48.0} & \better{21.0} \\
    {\it OpenAI, rerank} & \bad{64.0} & \good{\bf 72.6} & \bad{33.5} & \bad{\bf 37.4} & \bad{40.5} & {17.2} & \better{\bf 49.6} & \better{\bf 21.7} \\
    \hlrow {Gold} & {91.5} & {72.6} & {-} & {42.9} & {40.3} & {24.1} & {39.1} & {30.7} \\
    \bottomrule
  \end{tabular}
}
\vspace{-1mm}
\caption{Performance of retrieval-augmented code generation, with top retrieval and generation models. We bold-type the best RACG results. We test gpt-4o on SWE-bench to show non-trivial results than gpt-3.5-turbo. 
{Note that we exclude code canonical answer from the retrieval corpora for basic programming tasks.}}
\label{tab:baseline-rag-results}
\vspace{-3mm}
\end{table*}

\noindent \textbf{Basic programming problems} \quad
Most retrieved contexts help StarCoder2 generations. On MBPP, RACG even outperforms canonical setup by $15.6$--$17.8$. 
However, RACG does not improve DeepSeekCoder generations, which we observe is due to over-complicated and ungrammatically repetitive generations when with additional contexts. 
In comparison, GPT-3.5-turbo can effectively improve with added contexts, showing its better ability to leverage augmented contexts.

\noindent \textbf{Open-domain problems} \quad
The weaker StarCoder2 benefits from retrieved library documentation across all datasets, while DeepSeekCoder and GPT-3.5 show gains mainly on ODEX-hard problems. This aligns with findings from the canonical document setup, indicating that RACG is particularly effective for less popular libraries. Interestingly, despite relatively low NDCG@10 scores, the best-performing RACG combinations match their canonical results on ODEX-hard. 

\noindent \textbf{Repository-level problems} \quad
All models show improvements with retrieved code snippets on RepoEval, with RACG using strong retrievers like \texttt{openai-embeddings} performing on par with—or even surpassing—the canonical setup, likely due to the additional context provided to the models. On SWE-Bench, the best-performing combination, Retrieval-then-Rerank and GPT4o, yields a 21-point improvement over the no-retrieval baseline. However, there remains a 9-point gap compared to the gold setup, indicating room for improvement on the retrieval side, as reflected in the limited code retrieval performance shown in Table~\ref{tab:baseline-retrieval-results}.  



\begin{table*}[ht]
\small
\centering
\resizebox{1.00\textwidth}{!}{
  \begin{tabular}{ll|cccccc|c||cccccc|c}
    \toprule
    \multirow{2}{*}{\bf Model} & \multirow{2}{*}{\bf Retriever} & \multicolumn{7}{c||}{\bf HumanEval} &  \multicolumn{7}{c}{\bf ODEX} \\
    \cmidrule{3-16}
    {} & {} & {w/o} & \dlcell{Prog} & {Tut} & {Docs} & {SO} & {GitHub} & {All} & {w/o} & {Prog} & {Tut} & \dlcell{Docs} & {SO} & {GitHub} & {All} \\
    \midrule
    \multirow{3}{*}{StarCoder} & {BM25} & \multirow{3}{*}{31.7} & \dlcell{97.6} & {27.4} & {\bf 29.3} & \hlcell{32.9} & {30.5} & \hlcell{97.6} & \multirow{3}{*}{14.6} & \hlcell{18.2} & {13.4} & \dlcell{14.1} & {\bf 11.6} & \hlcell{15.9} & \hlcell{16.2} \\
    {} & {GIST} & {} & \dlcell{67.1} & \hlcell{\bf 34.8} & {26.7} & \hlcell{32.3} & \hlcell{\bf 32.9} & {69.1} & {} & {14.6} & \hlcell{\bf 15.7} & \dlcell{\bf 17.3} & {11.4} & \hlcell{15.5} & {\bf 17.1}  \\
    {} & {OpenAI} & {} & \dlcell{97.6} & {29.3} & {24.4} & \hlcell{\bf 36.0} & {31.1} & \hlcell{97.6} & {} & \hlcell{\bf 18.7} & {14.1} & \dlcell{15.9} & {10.9} & \hlcell{\bf 16.9} & \hlcell{15.3}\\
    \midrule
    {GPT-4o} & {OpenAI} & {75.6} & \dlcell{94.5} & \hlcell{90.2} & \hlcell{90.9} & \hlcell{91.5} & {84.8} & \hlcell{95.1} & {44.6} & \hlcell{\bf 49.2} & {44.2} & \dlcell{47.6} & {40.3} & {39.4} & {39.6}\\
    \bottomrule
  \end{tabular}
}
\vspace{-2mm}
\caption{\small{Comparing five retrieval sources on HumanEval and ODEX, using StarCoder2 (top) and GPT-4o (bottom).}}
\label{tab:humaneval-retrieval-sources}
\vspace{-3mm}
\end{table*}

\section{RACG with Open Retrieval}
\label{sec:4:retrieve-sources}
\vspace{-2mm}

Besides retrieving documents from the canonical source, we explore RACG with open retrieval from all sources (\S\ref{sec:2.2:retrieval-sources}) on three category-representative datasets: HumanEval, ODEX, and RepoEval.
We also study \textit{mixed} retrieval with documents from all sources, where we aggregate the top-1 documents from all five sources as augmented contexts.\footnote{We use the first 500 tokens of each document for all experiments in this section, which we show to be optimal in ablation studies (\S\ref{sec:4:retrieve-sources}), and satisfies all model context limits.}
We use the three top retrievers along with StarCoder2 and OpenAI retrieval with GPT-4o generation, to study open RACG with weak and strong LMs. 

\noindent \textbf{General programming: HumanEval} \quad
Among all sources, SO posts can improve the results by $1.8$--$4.3$, regardless of the choice of retrievers. Tutorials can improve results by $2.1$ only with the GIST retriever.
From manual examinations of the results, many retrieved posts and tutorials are \textit{about the same programming problem} as the HumanEval example, with code and detailed textual explanations, hence could hint or disclose the answer.
Other retrieval sources do not often contain relevant content thus do not bring improvements.
Surprisingly, generation with mixed documents performs as well as using the gold documents, suggesting that the model can \textit{discern and integrate the most useful content} from a mixture of texts.

\noindent \textbf{Open-domain: ODEX} \quad
Programming solutions are the most helpful source by bringing 3.8--4.3 gains
, even surpassing gains of canonical documentation. Notably, both GPT-4o and StarCoder using OpenAI retrieval from programming solutions, outperform their variants retrieving from documentation by 3.2 and 1.6 points. 
Although the retrieved content is only sometimes functionally relevant to the ODEX examples, they can \textit{exemplify the correct usage} of libraries such as \texttt{regex} in solutions and \texttt{requests} in GitHub files, thus guiding the generation to be more functionally correct.
Similar to HumanEval, GIST-large is particularly good at retrieving tutorials, while BM25 and OpenAI embeddings find higher-quality program solutions, indicating their respective domain advantages.

\noindent \textbf{Repository-level: RepoEval} \quad
Open sources are less useful than code snippets in the local repository. Understanding local code contexts is crucial and irreplaceable than external resources.
When using both local and open-source contexts (\textit{L+O}), models surpass the no-retrieval baseline, yet are still only comparable with \textit{Local}, suggesting more efforts and insights to benefit from both sources.

\begin{table}[ht]
\vspace{-1mm}
\small
\centering
\resizebox{0.50\textwidth}{!}{
  \begin{tabular}{l|ccccccc|cc}
    \toprule
    \textbf{Method} & {w/o} & \dlcell{Local} & {Prog} & {Tut} & {Docs} & {SO} & {GitHub} & {Open} & {L+O} \\
    \midrule
    \multicolumn{10}{c}{\it StarCoder2-7B} \\
    {BM25} & \multirow{3}{*}{26.5} & \dlcell{36.7} & {23.6} & {25.2} & {23.9} & {23.6} & {25.5} & {23.6} & \hlcell{31.4} \\
    {GIST} & {} & \dlcell{40.8} & {24.1} & {23.3} & {21.7} & {24.7} & {24.4} & {24.1} & \hlcell{41.8} \\
    {OpenAI} & {} & \dlcell{51.2} & {23.9} & {24.1} & {24.1} & {23.1} & {22.8} & {24.9} & \hlcell{\bf 50.9} \\
    \midrule
    \multicolumn{10}{c}{\it GPT-4o} \\
    {OpenAI} & {32.4} & \dlcell{62.2} & \hlcell{35.4} & {28.7} & {27.8} & {29.0} & {28.2} & {30.3} & \hlcell{54.2} \\
    \bottomrule
  \end{tabular}
}
\vspace{-2mm}
\caption{\small RACG with open retrieval on RepoEval.}
\label{tab:repoeval-retrieval-sources}
\vspace{-2mm}
\end{table}

\noindent \textbf{Exploring optimal chunking strategies} \quad
Adding many documents may exceed model context limits hence impairing RACG, we thus explore various chunking strategies to better integrate retrieval.
Compared to the no-chunking baseline, we study (i) post-retrieval chunking that takes the first N-tokens of each document, (ii) post-retrieval with reranking using BGE-reranker-base (\S\ref{sec:3.1:experiment-setup}) to find the most relevant N-token chunk from each document, and (iii) pre-retrieval chunking that chunks documents beforehand and retrieves N-token pieces directly.\footnote{We do not chunk programming solutions since they are typically short (average <200 tokens as in \autoref{tab:retrieval-sources}).} 

\begin{figure}[ht]
\centering
\vspace{-2mm}
\includegraphics[width=0.34\textwidth]{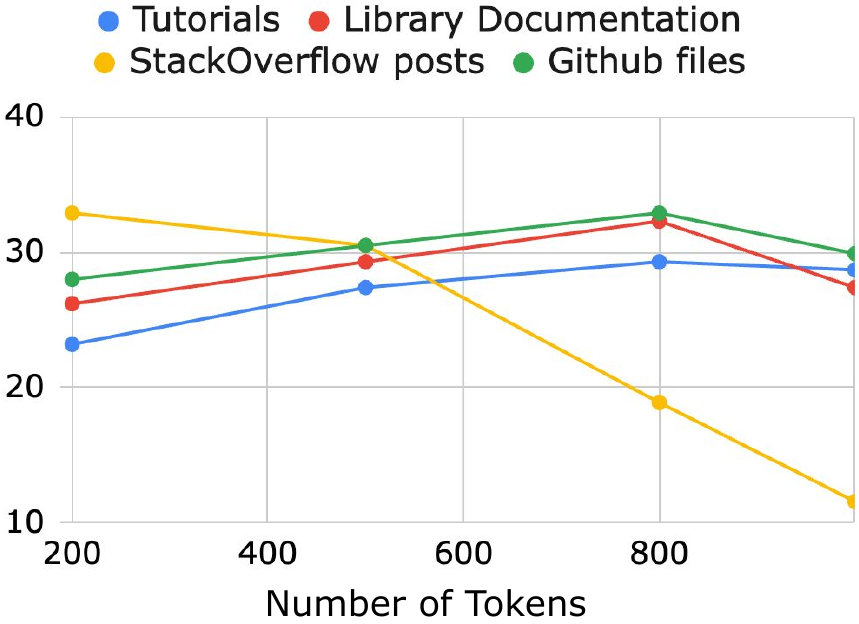}
\vspace{-2mm}
\caption{Performance with different chunking sizes.}
\label{fig:chunking-size}
\vspace{-2mm}
\end{figure}

\noindent We compare (i) using the first N-tokens for N from 200 to 1500 (\autoref{fig:chunking-size}). Most sources are best represented by the first 800 tokens except for SO posts.
However, we find (ii) reranking within this optimal range of 200--800 tokens greatly degrades the results, showing limited utility of current rerankers.
Lastly, (iii) pre-retrieval achieves the highest scores on almost all document sources (\autoref{tab:chunking-method}).

\begin{table}[ht]
\small 
\centering
\resizebox{0.44\textwidth}{!}{
  \begin{tabular}{l|cccc}
    \toprule
    \multicolumn{1}{c|}{\textbf{Method}} & {Tutorials} & {Docs} & {SO} & {GitHub} \\
    \midrule
    {Full text} & {~~6.7} & {17.7} & {28.0} & {~~3.7} \\
    {First chunk} & {27.4} & {29.3} & {30.5} & {\bf 30.5} \\
    {w/ reranking} & {~~9.1} & {~~9.1} & {14.0} & {13.4} \\
    {Pre-retrieval} & {\bf 31.1} & \hlcell {\bf 32.9} & \hlcell {\bf 33.5} & {29.3} \\
    \bottomrule
    \end{tabular}
}
\vspace{-2mm}
\caption{\small{Comparing chunking strategies on HumanEval.}}
\label{tab:chunking-method}
\vspace{-5mm}
\end{table}

\section{Related Work}
\textbf{Code generation} \quad Neural code generation has been a crucial task \citep{lu2021codexglue}, and increasingly strong code LMs have been created \citep{roziere2023code,li2023starcoder,guo2024deepseek,codegemma2024} to solve various tasks \citep{chen2021evaluating,lai2023ds,jimenez2024swebench}. 
However, most LMs generate code solely based on NL queries and model parametric knowledge, without using external programming sources (e.g., tutorials). 
To fill in this gap and allow systematic studies of RACG, we integrate various datasets and retrieval sources to build \datasetname.

\noindent \textbf{Retrieval augmented generation (RAG)} \quad
RAG has been widely used in knowledge-intensive tasks~\citep{lewis2020retrieval,guu2020retrieval},
however, mostly on text-centric tasks using general domain corpora such as Wikipedia~\cite{asai2024reliable}. 
Some works used programming context retrieved from repositories~\cite{ding2023crosscodeeval,yang2024swe} or documentations~\cite{zhou2023docprompting}, yet none of them considered RACG across varied coding tasks and knowledge sources. 
In text-centric tasks, unified benchmarks such as BEIR~\citep{thakur2021beir} and KILT~\citep{petroni2020kilt} aggregate retrieval and generation tasks and facilitate its progress~\cite{muennighoff2022mteb}. 
To similarly enable systematic studies of RACG across coding tasks and retrieval sources, we curate a unified benchmark and release its RACG codebase.

\vspace{-2mm}
\section{Conclusion}
\vspace{-2mm}
In this work, we propose \datasetname, a benchmark for retrieval-augmented code generation with various coding tasks and retrieval sources. With our experiments with top-performing retrieval and generation models, we show that retrieving external documents can greatly benefit code generation. However, current retrieval models struggle to find useful documents, and generation models have limited context capacity and RAG abilities, both leading to suboptimal RACG results. We hope \datasetname~can serve as a solid testbed to advance future endeavors in this direction.

\section*{Limitations}
We propose a new paradigm, retrieval-augmented code generation, equipped with a comprehensive benchmark \datasetname. However, as an initial exploration in this field, our work could be extended in task and language diversity, as well as model and methodological improvements.

We aggregate various existing code generation tasks, but many interesting scenarios such as code debugging remain under-explored. Meanwhile, we focus on coding tasks using Python programming language, but extrapolating to other languages may bring additional challenges. 

Meanwhile, for benchmarking purposes, we mostly experimented with vanilla retrieval, reranking, and generation methods, but better backbone models and advanced methods for each RACG component are yet fully explored. Our results may not represent all model behaviors, and we encourage future works to build methods that break certain limitations we observe in current systems.

\section*{Acknowledgment}
We thank Shuyan Zhou and Xinran Zhao for the helpful discussions in the early stage of this project; Saujas Vaduguru, Jing Yu Koh, Alex Xie, and Andy Liu for providing valuable feedback for the draft. 

\bibliography{custom}

\begin{thebibliography}{48}
\providecommand{\natexlab}[1]{#1}

\bibitem[{Asai et~al.(2023)Asai, Schick, Lewis, Chen, Izacard, Riedel, Hajishirzi, and Yih}]{asai-etal-2023-task}
Akari Asai, Timo Schick, Patrick Lewis, Xilun Chen, Gautier Izacard, Sebastian Riedel, Hannaneh Hajishirzi, and Wen-tau Yih. 2023.
\newblock \href {https://doi.org/10.18653/v1/2023.findings-acl.225} {Task-aware retrieval with instructions}.
\newblock In \emph{Findings of the Association for Computational Linguistics: ACL 2023}, pages 3650--3675, Toronto, Canada. Association for Computational Linguistics.

\bibitem[{Asai et~al.(2024)Asai, Zhong, Chen, Koh, Zettlemoyer, Hajishirzi, and Yih}]{asai2024reliable}
Akari Asai, Zexuan Zhong, Danqi Chen, Pang~Wei Koh, Luke Zettlemoyer, Hannaneh Hajishirzi, and Wen-tau Yih. 2024.
\newblock Reliable, adaptable, and attributable language models with retrieval.
\newblock \emph{arXiv preprint arXiv:2403.03187}.

\bibitem[{Austin et~al.(2021)Austin, Odena, Nye, Bosma, Michalewski, Dohan, Jiang, Cai, Terry, Le et~al.}]{austin2021program}
Jacob Austin, Augustus Odena, Maxwell Nye, Maarten Bosma, Henryk Michalewski, David Dohan, Ellen Jiang, Carrie Cai, Michael Terry, Quoc Le, et~al. 2021.
\newblock Program synthesis with large language models.
\newblock \emph{arXiv preprint arXiv:2108.07732}.

\bibitem[{Brown et~al.(2020)Brown, Mann, Ryder, Subbiah, Kaplan, Dhariwal, Neelakantan, Shyam, Sastry, Askell, Agarwal, Herbert-Voss, Krueger, Henighan, Child, Ramesh, Ziegler, Wu, Winter, Hesse, Chen, Sigler, Litwin, Gray, Chess, Clark, Berner, McCandlish, Radford, Sutskever, and Amodei}]{Brown2020LanguageMA}
Tom~B. Brown, Benjamin Mann, Nick Ryder, Melanie Subbiah, Jared Kaplan, Prafulla Dhariwal, Arvind Neelakantan, Pranav Shyam, Girish Sastry, Amanda Askell, Sandhini Agarwal, Ariel Herbert-Voss, Gretchen Krueger, Tom Henighan, Rewon Child, Aditya Ramesh, Daniel~M. Ziegler, Jeff Wu, Clemens Winter, Christopher Hesse, Mark Chen, Eric Sigler, Mateusz Litwin, Scott Gray, Benjamin Chess, Jack Clark, Christopher Berner, Sam McCandlish, Alec Radford, Ilya Sutskever, and Dario Amodei. 2020.
\newblock \href {https://api.semanticscholar.org/CorpusID:218971783} {Language models are few-shot learners}.
\newblock \emph{ArXiv}, abs/2005.14165.

\bibitem[{Chen et~al.(2021)Chen, Tworek, Jun, Yuan, Pinto, Kaplan, Edwards, Burda, Joseph, Brockman et~al.}]{chen2021evaluating}
Mark Chen, Jerry Tworek, Heewoo Jun, Qiming Yuan, Henrique Ponde de~Oliveira Pinto, Jared Kaplan, Harri Edwards, Yuri Burda, Nicholas Joseph, Greg Brockman, et~al. 2021.
\newblock Evaluating large language models trained on code.
\newblock \emph{arXiv preprint arXiv:2107.03374}.

\bibitem[{CohereAI(2024)}]{command-r}
CohereAI. 2024.
\newblock \href {https://docs.cohere.com/docs/command-r} {Command r}.

\bibitem[{Computer(2023)}]{together2023redpajama}
Together Computer. 2023.
\newblock \href {https://github.com/togethercomputer/RedPajama-Data} {Redpajama: An open source recipe to reproduce llama training dataset}.

\bibitem[{Ding et~al.(2023)Ding, Wang, Ahmad, Ding, Tan, Jain, Ramanathan, Nallapati, Bhatia, Roth, and Xiang}]{ding2023crosscodeeval}
Yangruibo Ding, Zijian Wang, Wasi~Uddin Ahmad, Hantian Ding, Ming Tan, Nihal Jain, Murali~Krishna Ramanathan, Ramesh Nallapati, Parminder Bhatia, Dan Roth, and Bing Xiang. 2023.
\newblock \href {https://openreview.net/forum?id=wgDcbBMSfh} {Crosscodeeval: A diverse and multilingual benchmark for cross-file code completion}.
\newblock In \emph{Thirty-seventh Conference on Neural Information Processing Systems Datasets and Benchmarks Track}.

\bibitem[{G{\"u}nther et~al.(2023)G{\"u}nther, Ong, Mohr, Abdessalem, Abel, Akram, Guzman, Mastrapas, Sturua, Wang et~al.}]{gunther2023jina}
Michael G{\"u}nther, Jackmin Ong, Isabelle Mohr, Alaeddine Abdessalem, Tanguy Abel, Mohammad~Kalim Akram, Susana Guzman, Georgios Mastrapas, Saba Sturua, Bo~Wang, et~al. 2023.
\newblock Jina embeddings 2: 8192-token general-purpose text embeddings for long documents.
\newblock \emph{arXiv preprint arXiv:2310.19923}.

\bibitem[{Guo et~al.(2024)Guo, Zhu, Yang, Xie, Dong, Zhang, Chen, Bi, Wu, Li et~al.}]{guo2024deepseek}
Daya Guo, Qihao Zhu, Dejian Yang, Zhenda Xie, Kai Dong, Wentao Zhang, Guanting Chen, Xiao Bi, Y~Wu, YK~Li, et~al. 2024.
\newblock Deepseek-coder: When the large language model meets programming--the rise of code intelligence.
\newblock \emph{arXiv preprint arXiv:2401.14196}.

\bibitem[{Guu et~al.(2020)Guu, Lee, Tung, Pasupat, and Chang}]{guu2020retrieval}
Kelvin Guu, Kenton Lee, Zora Tung, Panupong Pasupat, and Mingwei Chang. 2020.
\newblock Retrieval augmented language model pre-training.
\newblock In \emph{International conference on machine learning}, pages 3929--3938. PMLR.

\bibitem[{Izacard et~al.(2022{\natexlab{a}})Izacard, Caron, Hosseini, Riedel, Bojanowski, Joulin, and Grave}]{izacard2022unsupervised}
Gautier Izacard, Mathilde Caron, Lucas Hosseini, Sebastian Riedel, Piotr Bojanowski, Armand Joulin, and Edouard Grave. 2022{\natexlab{a}}.
\newblock \href {https://openreview.net/forum?id=jKN1pXi7b0} {Unsupervised dense information retrieval with contrastive learning}.
\newblock \emph{Transactions on Machine Learning Research}.

\bibitem[{Izacard et~al.(2022{\natexlab{b}})Izacard, Lewis, Lomeli, Hosseini, Petroni, Schick, Yu, Joulin, Riedel, and Grave}]{Izacard2022FewshotLW}
Gautier Izacard, Patrick Lewis, Maria Lomeli, Lucas Hosseini, Fabio Petroni, Timo Schick, Jane~A. Yu, Armand Joulin, Sebastian Riedel, and Edouard Grave. 2022{\natexlab{b}}.
\newblock \href {https://api.semanticscholar.org/CorpusID:251371732} {Few-shot learning with retrieval augmented language models}.
\newblock \emph{ArXiv}, abs/2208.03299.

\bibitem[{Jain et~al.(2024)Jain, Han, Gu, Li, Yan, Zhang, Wang, Solar-Lezama, Sen, and Stoica}]{jain2024livecodebench}
Naman Jain, King Han, Alex Gu, Wen-Ding Li, Fanjia Yan, Tianjun Zhang, Sida Wang, Armando Solar-Lezama, Koushik Sen, and Ion Stoica. 2024.
\newblock Livecodebench: Holistic and contamination free evaluation of large language models for code.
\newblock \emph{arXiv preprint arXiv:2403.07974}.

\bibitem[{Jimenez et~al.(2024)Jimenez, Yang, Wettig, Yao, Pei, Press, and Narasimhan}]{jimenez2024swebench}
Carlos~E Jimenez, John Yang, Alexander Wettig, Shunyu Yao, Kexin Pei, Ofir Press, and Karthik~R Narasimhan. 2024.
\newblock \href {https://openreview.net/forum?id=VTF8yNQM66} {{SWE}-bench: Can language models resolve real-world github issues?}
\newblock In \emph{The Twelfth International Conference on Learning Representations}.

\bibitem[{Kandpal et~al.(2023)Kandpal, Deng, Roberts, Wallace, and Raffel}]{kandpal2023large}
Nikhil Kandpal, Haikang Deng, Adam Roberts, Eric Wallace, and Colin Raffel. 2023.
\newblock Large language models struggle to learn long-tail knowledge.
\newblock In \emph{International Conference on Machine Learning}, pages 15696--15707. PMLR.

\bibitem[{Lai et~al.(2023)Lai, Li, Wang, Zhang, Zhong, Zettlemoyer, Yih, Fried, Wang, and Yu}]{lai2023ds}
Yuhang Lai, Chengxi Li, Yiming Wang, Tianyi Zhang, Ruiqi Zhong, Luke Zettlemoyer, Wen-tau Yih, Daniel Fried, Sida Wang, and Tao Yu. 2023.
\newblock Ds-1000: a natural and reliable benchmark for data science code generation.
\newblock In \emph{Proceedings of the 40th International Conference on Machine Learning}, ICML'23. JMLR.org.

\bibitem[{Lewis et~al.(2020)Lewis, Perez, Piktus, Petroni, Karpukhin, Goyal, K\"{u}ttler, Lewis, Yih, Rockt\"{a}schel, Riedel, and Kiela}]{lewis2020retrieval}
Patrick Lewis, Ethan Perez, Aleksandra Piktus, Fabio Petroni, Vladimir Karpukhin, Naman Goyal, Heinrich K\"{u}ttler, Mike Lewis, Wen-tau Yih, Tim Rockt\"{a}schel, Sebastian Riedel, and Douwe Kiela. 2020.
\newblock \href {https://proceedings.neurips.cc/paper_files/paper/2020/file/6b493230205f780e1bc26945df7481e5-Paper.pdf} {Retrieval-augmented generation for knowledge-intensive nlp tasks}.
\newblock In \emph{Advances in Neural Information Processing Systems}, volume~33, pages 9459--9474. Curran Associates, Inc.

\bibitem[{Li et~al.(2023)Li, allal, Zi, Muennighoff, Kocetkov, Mou, Marone, Akiki, LI, Chim, Liu, Zheltonozhskii, Zhuo, Wang, Dehaene, Lamy-Poirier, Monteiro, Gontier, Yee, Umapathi, Zhu, Lipkin, Oblokulov, Wang, Murthy, Stillerman, Patel, Abulkhanov, Zocca, Dey, Zhang, Bhattacharyya, Yu, Luccioni, Villegas, Zhdanov, Lee, Timor, Ding, Schlesinger, Schoelkopf, Ebert, Dao, Mishra, Gu, Anderson, Dolan-Gavitt, Contractor, Reddy, Fried, Bahdanau, Jernite, Ferrandis, Hughes, Wolf, Guha, Werra, and de~Vries}]{li2023starcoder}
Raymond Li, Loubna~Ben allal, Yangtian Zi, Niklas Muennighoff, Denis Kocetkov, Chenghao Mou, Marc Marone, Christopher Akiki, Jia LI, Jenny Chim, Qian Liu, Evgenii Zheltonozhskii, Terry~Yue Zhuo, Thomas Wang, Olivier Dehaene, Joel Lamy-Poirier, Joao Monteiro, Nicolas Gontier, Ming-Ho Yee, Logesh~Kumar Umapathi, Jian Zhu, Ben Lipkin, Muhtasham Oblokulov, Zhiruo Wang, Rudra Murthy, Jason~T Stillerman, Siva~Sankalp Patel, Dmitry Abulkhanov, Marco Zocca, Manan Dey, Zhihan Zhang, Urvashi Bhattacharyya, Wenhao Yu, Sasha Luccioni, Paulo Villegas, Fedor Zhdanov, Tony Lee, Nadav Timor, Jennifer Ding, Claire~S Schlesinger, Hailey Schoelkopf, Jan Ebert, Tri Dao, Mayank Mishra, Alex Gu, Carolyn~Jane Anderson, Brendan Dolan-Gavitt, Danish Contractor, Siva Reddy, Daniel Fried, Dzmitry Bahdanau, Yacine Jernite, Carlos~Mu{\~n}oz Ferrandis, Sean Hughes, Thomas Wolf, Arjun Guha, Leandro~Von Werra, and Harm de~Vries. 2023.
\newblock \href {https://openreview.net/forum?id=KoFOg41haE} {Starcoder: may the source be with you!}
\newblock \emph{Transactions on Machine Learning Research}.
\newblock Reproducibility Certification.

\bibitem[{Li et~al.(2022)Li, Choi, Chung, Kushman, Schrittwieser, Leblond, Eccles, Keeling, Gimeno, Lago, Hubert, Choy, de~Masson~d’Autume, Babuschkin, Chen, Huang, Welbl, Gowal, Cherepanov, Molloy, Mankowitz, Robson, Kohli, de~Freitas, Kavukcuoglu, and Vinyals}]{yujia2022competition}
Yujia Li, David Choi, Junyoung Chung, Nate Kushman, Julian Schrittwieser, Rémi Leblond, Tom Eccles, James Keeling, Felix Gimeno, Agustin~Dal Lago, Thomas Hubert, Peter Choy, Cyprien de~Masson~d’Autume, Igor Babuschkin, Xinyun Chen, Po-Sen Huang, Johannes Welbl, Sven Gowal, Alexey Cherepanov, James Molloy, Daniel~J. Mankowitz, Esme~Sutherland Robson, Pushmeet Kohli, Nando de~Freitas, Koray Kavukcuoglu, and Oriol Vinyals. 2022.
\newblock \href {https://doi.org/10.1126/science.abq1158} {Competition-level code generation with alphacode}.
\newblock \emph{Science}, 378(6624):1092--1097.

\bibitem[{Lin et~al.(2021)Lin, Ma, Lin, Yang, Pradeep, and Nogueira}]{Lin_etal_SIGIR2021_Pyserini}
Jimmy Lin, Xueguang Ma, Sheng-Chieh Lin, Jheng-Hong Yang, Ronak Pradeep, and Rodrigo Nogueira. 2021.
\newblock \href {https://dl.acm.org/doi/10.1145/3404835.3463238} {{Pyserini}: A {Python} toolkit for reproducible information retrieval research with sparse and dense representations}.
\newblock In \emph{Proceedings of the 44th Annual International ACM SIGIR Conference on Research and Development in Information Retrieval (SIGIR 2021)}.

\bibitem[{Liu et~al.(2023)Liu, Xia, Wang, and Zhang}]{evalplus}
Jiawei Liu, Chunqiu~Steven Xia, Yuyao Wang, and Lingming Zhang. 2023.
\newblock \href {https://openreview.net/forum?id=1qvx610Cu7} {Is your code generated by chat{GPT} really correct? rigorous evaluation of large language models for code generation}.
\newblock In \emph{Thirty-seventh Conference on Neural Information Processing Systems}.

\bibitem[{Lozhkov et~al.(2024)Lozhkov, Li, Allal, Cassano, Lamy-Poirier, Tazi, Tang, Pykhtar, Liu, Wei et~al.}]{lozhkov2024starcoder}
Anton Lozhkov, Raymond Li, Loubna~Ben Allal, Federico Cassano, Joel Lamy-Poirier, Nouamane Tazi, Ao~Tang, Dmytro Pykhtar, Jiawei Liu, Yuxiang Wei, et~al. 2024.
\newblock Starcoder 2 and the stack v2: The next generation.
\newblock \emph{arXiv preprint arXiv:2402.19173}.

\bibitem[{Lu et~al.(2021)Lu, Guo, Ren, Huang, Svyatkovskiy, Blanco, Clement, Drain, Jiang, Tang, Li, Zhou, Shou, Zhou, Tufano, Gong, Zhou, Duan, Sundaresan, Deng, Fu, and Liu}]{lu2021codexglue}
Shuai Lu, Daya Guo, Shuo Ren, Junjie Huang, Alexey Svyatkovskiy, Ambrosio Blanco, Colin~B. Clement, Dawn Drain, Daxin Jiang, Duyu Tang, Ge~Li, Lidong Zhou, Linjun Shou, Long Zhou, Michele Tufano, Ming Gong, Ming Zhou, Nan Duan, Neel Sundaresan, Shao~Kun Deng, Shengyu Fu, and Shujie Liu. 2021.
\newblock Codexglue: {A} machine learning benchmark dataset for code understanding and generation.
\newblock \emph{CoRR}, abs/2102.04664.

\bibitem[{Mallen et~al.(2023)Mallen, Asai, Zhong, Das, Khashabi, and Hajishirzi}]{mallen-etal-2023-trust}
Alex Mallen, Akari Asai, Victor Zhong, Rajarshi Das, Daniel Khashabi, and Hannaneh Hajishirzi. 2023.
\newblock \href {https://doi.org/10.18653/v1/2023.acl-long.546} {When not to trust language models: Investigating effectiveness of parametric and non-parametric memories}.
\newblock In \emph{Proceedings of the 61st Annual Meeting of the Association for Computational Linguistics (Volume 1: Long Papers)}, pages 9802--9822, Toronto, Canada. Association for Computational Linguistics.

\bibitem[{Meng et~al.(2024)Meng, Liu, Joty, Xiong, Zhou, and Yavuz}]{SFRAIResearch2024}
Rui Meng, Ye~Liu, Shafiq~Rayhan Joty, Caiming Xiong, Yingbo Zhou, and Semih Yavuz. 2024.
\newblock \href {https://blog.salesforceairesearch.com/sfr-embedded-mistral/} {Sfr-embedding-mistral:enhance text retrieval with transfer learning}.

\bibitem[{Meta(2024)}]{introducing2024}
Meta. 2024.
\newblock \href {https://ai.meta.com/blog/meta-llama-3/} {Introducing meta llama 3: The most capable openly available llm to date}.

\bibitem[{Muennighoff et~al.(2022)Muennighoff, Tazi, Magne, and Reimers}]{muennighoff2022mteb}
Niklas Muennighoff, Nouamane Tazi, Lo{\"\i}c Magne, and Nils Reimers. 2022.
\newblock Mteb: Massive text embedding benchmark.
\newblock \emph{arXiv preprint arXiv:2210.07316}.

\bibitem[{Overwijk et~al.(2022)Overwijk, Xiong, Liu, VandenBerg, and Callan}]{overwijk2022clueweb22}
Arnold Overwijk, Chenyan Xiong, Xiao Liu, Cameron VandenBerg, and Jamie Callan. 2022.
\newblock Clueweb22: 10 billion web documents with visual and semantic information.
\newblock \emph{arXiv preprint arXiv:2211.15848}.

\bibitem[{Petroni et~al.(2020)Petroni, Piktus, Fan, Lewis, Yazdani, De~Cao, Thorne, Jernite, Karpukhin, Maillard et~al.}]{petroni2020kilt}
Fabio Petroni, Aleksandra Piktus, Angela Fan, Patrick Lewis, Majid Yazdani, Nicola De~Cao, James Thorne, Yacine Jernite, Vladimir Karpukhin, Jean Maillard, et~al. 2020.
\newblock Kilt: a benchmark for knowledge intensive language tasks.
\newblock \emph{arXiv preprint arXiv:2009.02252}.

\bibitem[{Reimers and Gurevych(2019)}]{reimers-2019-sentence-bert}
Nils Reimers and Iryna Gurevych. 2019.
\newblock \href {https://arxiv.org/abs/1908.10084} {Sentence-bert: Sentence embeddings using siamese bert-networks}.
\newblock In \emph{Proceedings of the 2019 Conference on Empirical Methods in Natural Language Processing}. Association for Computational Linguistics.

\bibitem[{Robertson and Zaragoza(2009)}]{Robertson2009ThePR}
Stephen~E. Robertson and Hugo Zaragoza. 2009.
\newblock \href {https://api.semanticscholar.org/CorpusID:207178704} {The probabilistic relevance framework: Bm25 and beyond}.
\newblock \emph{Found. Trends Inf. Retr.}, 3:333--389.

\bibitem[{Roziere et~al.(2023)Roziere, Gehring, Gloeckle, Sootla, Gat, Tan, Adi, Liu, Remez, Rapin et~al.}]{roziere2023code}
Baptiste Roziere, Jonas Gehring, Fabian Gloeckle, Sten Sootla, Itai Gat, Xiaoqing~Ellen Tan, Yossi Adi, Jingyu Liu, Tal Remez, J{\'e}r{\'e}my Rapin, et~al. 2023.
\newblock Code llama: Open foundation models for code.
\newblock \emph{arXiv preprint arXiv:2308.12950}.

\bibitem[{Solatorio(2024)}]{solatorio2024gistembed}
Aivin~V. Solatorio. 2024.
\newblock \href {https://arxiv.org/abs/2402.16829} {Gistembed: Guided in-sample selection of training negatives for text embedding fine-tuning}.

\bibitem[{Su et~al.(2024)Su, Jiang, Lai, Wu, Shi, Liu, Liu, and Yu}]{su2024arks}
Hongjin Su, Shuyang Jiang, Yuhang Lai, Haoyuan Wu, Boao Shi, Che Liu, Qian Liu, and Tao Yu. 2024.
\newblock Arks: Active retrieval in knowledge soup for code generation.
\newblock \emph{arXiv preprint arXiv:2402.12317}.

\bibitem[{Su et~al.(2023)Su, Shi, Kasai, Wang, Hu, Ostendorf, Yih, Smith, Zettlemoyer, and Yu}]{su-etal-2023-one}
Hongjin Su, Weijia Shi, Jungo Kasai, Yizhong Wang, Yushi Hu, Mari Ostendorf, Wen-tau Yih, Noah~A. Smith, Luke Zettlemoyer, and Tao Yu. 2023.
\newblock \href {https://doi.org/10.18653/v1/2023.findings-acl.71} {One embedder, any task: Instruction-finetuned text embeddings}.
\newblock In \emph{Findings of the Association for Computational Linguistics: ACL 2023}, pages 1102--1121, Toronto, Canada. Association for Computational Linguistics.

\bibitem[{Team(2024)}]{codegemma2024}
CodeGemma Team. 2024.
\newblock \href {https://storage.googleapis.com/deepmind-media/gemma/codegemma_report.pdf} {Codegemma: Open code models based on gemma}.

\bibitem[{Thakur et~al.(2021)Thakur, Reimers, R{\"u}ckl{\'e}, Srivastava, and Gurevych}]{thakur2021beir}
Nandan Thakur, Nils Reimers, Andreas R{\"u}ckl{\'e}, Abhishek Srivastava, and Iryna Gurevych. 2021.
\newblock \href {https://openreview.net/forum?id=wCu6T5xFjeJ} {{BEIR}: A heterogeneous benchmark for zero-shot evaluation of information retrieval models}.
\newblock In \emph{Thirty-fifth Conference on Neural Information Processing Systems Datasets and Benchmarks Track (Round 2)}.

\bibitem[{VoyageAI(2024)}]{voyage2024voyage}
VoyageAI. 2024.
\newblock \href {https://blog.voyageai.com/2024/01/23/voyage-code-2-elevate-your-code-retrieval/} {voyage-code-2: Elevate your code retrieval}.

\bibitem[{Wang et~al.(2024)Wang, Li, Song, Xu, Tang, Zhuge, Pan, Song, Li, Singh et~al.}]{wang2024opendevin}
Xingyao Wang, Boxuan Li, Yufan Song, Frank~F Xu, Xiangru Tang, Mingchen Zhuge, Jiayi Pan, Yueqi Song, Bowen Li, Jaskirat Singh, et~al. 2024.
\newblock Opendevin: An open platform for ai software developers as generalist agents.
\newblock \emph{arXiv preprint arXiv:2407.16741}.

\bibitem[{Wang et~al.(2023{\natexlab{a}})Wang, Araki, Jiang, Parvez, and Neubig}]{wang2023learning}
Zhiruo Wang, Jun Araki, Zhengbao Jiang, Md~Rizwan Parvez, and Graham Neubig. 2023{\natexlab{a}}.
\newblock Learning to filter context for retrieval-augmented generation.
\newblock \emph{arXiv preprint arXiv:2311.08377}.

\bibitem[{Wang et~al.(2023{\natexlab{b}})Wang, Zhou, Fried, and Neubig}]{wang2023execution}
Zhiruo Wang, Shuyan Zhou, Daniel Fried, and Graham Neubig. 2023{\natexlab{b}}.
\newblock \href {https://doi.org/10.18653/v1/2023.findings-emnlp.89} {Execution-based evaluation for open-domain code generation}.
\newblock In \emph{Findings of the Association for Computational Linguistics: EMNLP 2023}. Association for Computational Linguistics.

\bibitem[{Xia et~al.(2024)Xia, Deng, Dunn, and Zhang}]{xia2024agentless}
Chunqiu~Steven Xia, Yinlin Deng, Soren Dunn, and Lingming Zhang. 2024.
\newblock Agentless: Demystifying llm-based software engineering agents.
\newblock \emph{arXiv preprint arXiv:2407.01489}.

\bibitem[{Xiao et~al.(2023)Xiao, Liu, Zhang, and Muennighoff}]{bge_embedding}
Shitao Xiao, Zheng Liu, Peitian Zhang, and Niklas Muennighoff. 2023.
\newblock \href {https://arxiv.org/abs/2309.07597} {C-pack: Packaged resources to advance general chinese embedding}.
\newblock \emph{arXiv}.

\bibitem[{Yang et~al.(2024)Yang, Jimenez, Wettig, Lieret, Yao, Narasimhan, and Press}]{yang2024swe}
John Yang, Carlos~E Jimenez, Alexander Wettig, Kilian Lieret, Shunyu Yao, Karthik Narasimhan, and Ofir Press. 2024.
\newblock Swe-agent: Agent-computer interfaces enable automated software engineering.
\newblock \emph{arXiv preprint arXiv:2405.15793}.

\bibitem[{Zhang et~al.(2024)Zhang, Ahmad, Tan, Ding, Nallapati, Roth, Ma, and Xiang}]{zhang2024code}
Dejiao Zhang, Wasi Ahmad, Ming Tan, Hantian Ding, Ramesh Nallapati, Dan Roth, Xiaofei Ma, and Bing Xiang. 2024.
\newblock Code representation learning at scale.
\newblock \emph{arXiv preprint arXiv:2402.01935}.

\bibitem[{Zhang et~al.(2023)Zhang, Chen, Zhang, Keung, Liu, Zan, Mao, Lou, and Chen}]{zhang2023repocoder}
Fengji Zhang, Bei Chen, Yue Zhang, Jacky Keung, Jin Liu, Daoguang Zan, Yi~Mao, Jian-Guang Lou, and Weizhu Chen. 2023.
\newblock \href {https://aclanthology.org/2023.emnlp-main.151} {Repocoder: Repository-level code completion through iterative retrieval and generation}.
\newblock In \emph{Proceedings of the 2023 Conference on Empirical Methods in Natural Language Processing}. Association for Computational Linguistics.

\bibitem[{Zhou et~al.(2023)Zhou, Alon, Xu, Jiang, and Neubig}]{zhou2023docprompting}
Shuyan Zhou, Uri Alon, Frank~F. Xu, Zhengbao Jiang, and Graham Neubig. 2023.
\newblock \href {https://openreview.net/forum?id=ZTCxT2t2Ru} {Docprompting: Generating code by retrieving the docs}.
\newblock In \emph{The Eleventh International Conference on Learning Representations}.

\end{thebibliography}

\clearpage
\appendix

\section{Example Illustrations}
\label{app:experiment-details}
\vspace{-2mm}

\subsection{Example with Canonical Documents}

\begin{figure*}[ht]
    \centering
    \includegraphics[width=\textwidth]{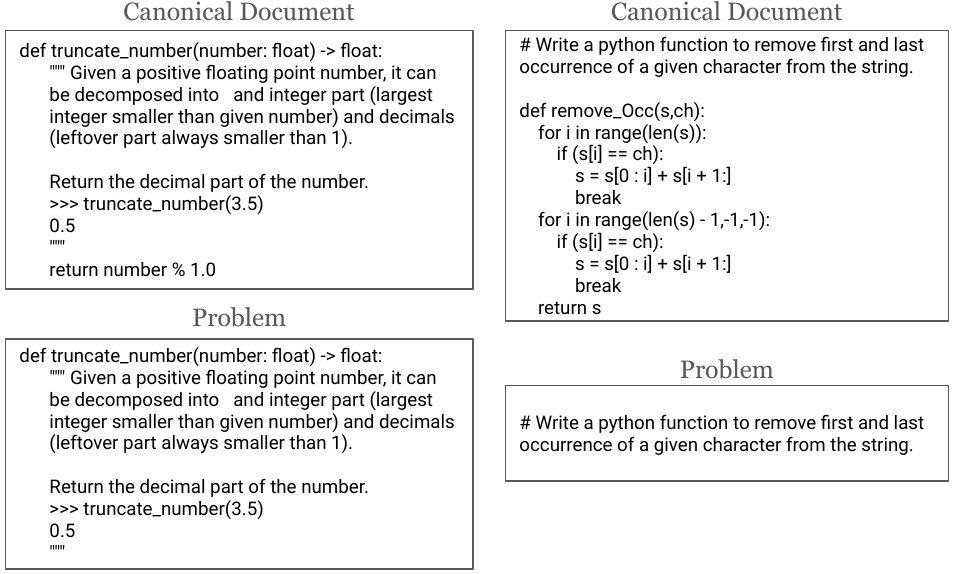}
    \caption{HumanEval (left) and MBPP (right) examples with annotated canonical solutions.}
    \label{fig:general-programming-examples}
\vspace{-4mm}
\end{figure*}

\begin{figure*}[ht]
    \centering
    \includegraphics[width=\textwidth]{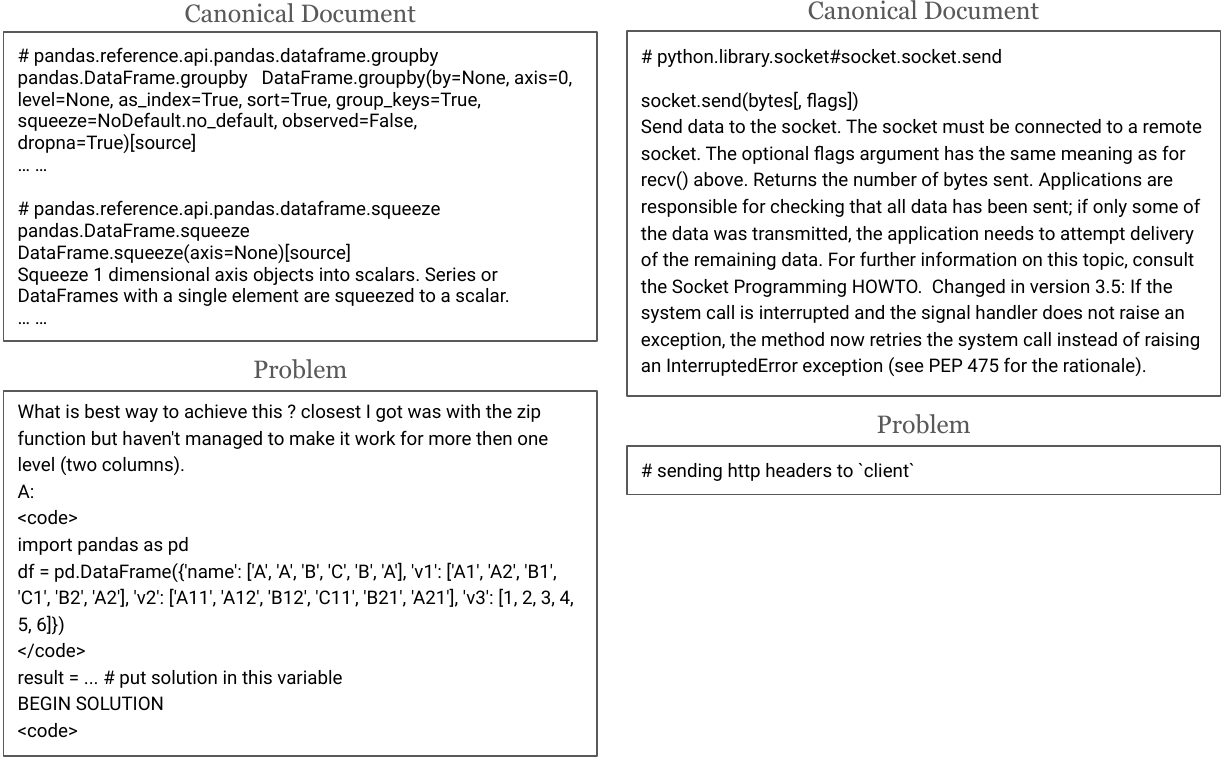}
    \caption{DS-1000 (left) and ODEX (right) examples with annotated canonical library documentation.}
    \label{fig:open-domain-examples}
\end{figure*}

To present our canonical document annotation (\S\ref{sec:2.3:canonical-document}) more concretely, we illustrate examples with their annotated canonical documents. 
\autoref{fig:general-programming-examples} shows the general-programming examples, with one HumanEval and one MBPP example, respectively.
\autoref{fig:open-domain-examples} shows two open-domain coding examples with canonical library documentation from DS-1000 and ODEX, respectively.

\subsection{RACG with Helpful and Distracting Documents}
Beyond the numerical numbers reported in experiment sections, here we provide some concrete examples that: (i) benefit from RACG when relevant documents are retrieved, and (ii) distracted by irrelevant documents retrieved hence results in degraded performance.

\begin{figure*}[ht]
    \centering
    \includegraphics[width=\textwidth]{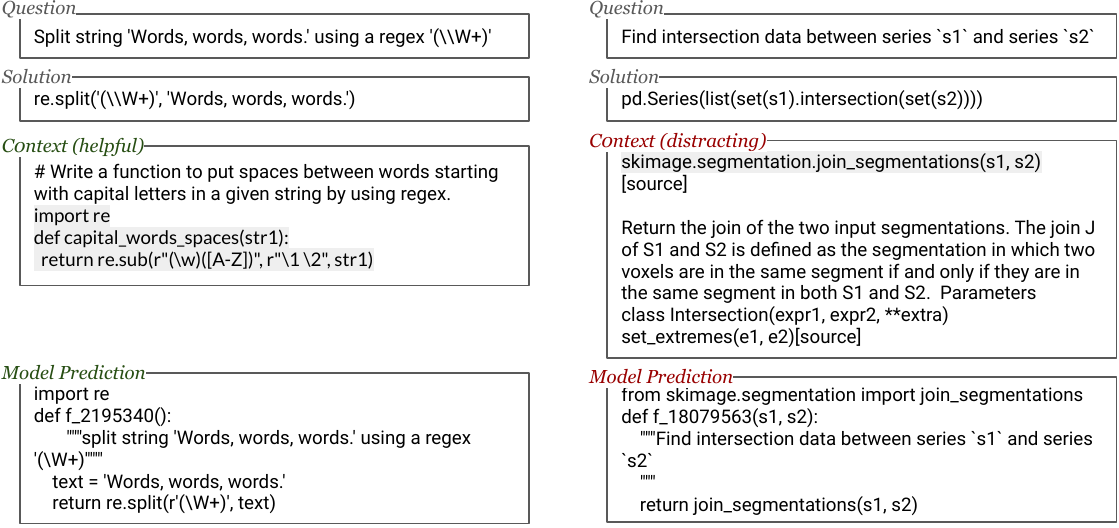}
    \caption{RACG helps with relevant contexts (left) and hurts with distracting contexts (right).}
    \label{fig:racg-examples}
\end{figure*}
\section{Additional Details about Retrieval Efficiency}
\label{app:retrieval-cost}
For open access models, we use the same single A100 GPU with 80 GRAM, with a batch size of 64 for GIST base and large, and 8 for SFR-Mistral. 
For proprietary models, we estimate their efficiency using a batch size of 64. We then average the time for each batch for each query and document. For Voyage-code, we apply a ``dynamic-batching'' technique that make sure the total tokens in the batch won't exceed the token limit. For both open and proprietary models, we define the search efficiency as the time it takes to embed individual query and the time to calculate similarities. Note that the time for both can be optimized by tokenizing all documents and all queries, then taking the dot product. 
The actual runtime for API models varies for each organization with different rate limits and the batch size. 
For this experiment, we set the maximum context length to match the maximum length of the original models. This notably increases the encoding latency of SFR Mixtral, which has a longer maximum context window size than smaller embedding models.

\section{Result Reproduction}
\label{app:reproduce-results}
\vspace{-2mm}

In \autoref{tab:baseline-generation-results} in \S\ref{sec:3:baseline}, we are able to reproduce most results reported in the original papers, but with minor variances. Here we explain the differences in implementation and (potential) reasons that lead to these small performance variances.

\noindent \textbf{Our approach} \quad
To keep a fair comparison among all models, we use the same prompt for each dataset when evaluating all models. Meanwhile, we use zero-shot prompts without any additional instructions, i.e., only input the original problem description of the example, to prevent unknown effects on the model performance when using different instructions and/or in-context examples.

According to this setup, we next describe the differences in prompts used by the original works and how they may affect the results.

\noindent \textbf{StarCoder2} \quad
The StarCoder2 technical report \citep{lozhkov2024starcoder} reported results on the HumanEval, MBPP, and DS-1000 datasets. 
On HumanEval, our reproduced results (31.7) is slightly lower than their number (35.4), possibly because the original paper additionally input the test cases as additional information in the prompt, whereas in our basic NL-to-code setup, no test cases are provided. This additional information may cause their results to be higher.

On MBPP dataset, they adopt a subset of MBPP, i.e., 399 out of 427 examples that have additional test cases populated by \citet{evalplus}. In contrast, we evaluate on the entire dataset, which is likely to cause the variance in results.

On DS-1000, the original paper samples 40 generations and report the pass@1 rate, while we only generate one program with greedy decoding. This difference in decoding strategy may cause slight variance in the results.

\noindent \textbf{CodeGemma} \quad
The CodeGemma technical report \citep{codegemma2024} reported results on HumanEval and MBPP datasets, but does not provide any details about the instructions, few-shot examples, or other parts of the prompt that they use. We were able to roughly reproduce their reported results, but with 3-5 points less in pass@1.

\noindent \textbf{CodeLlama} \quad
The CodeLlama technical report \citep{roziere2023code} reports results on HumanEval and MBPP datasets. We were able to perfectly reproduce their results on the HumanEval dataset under the zero-shot setting. However, for MBPP experiments, they use 3-shot prompting, which could potentially explain that our zero-shot results are 4 points lower in pass@1. 

\noindent \textbf{DeepSeekCoder} \quad
The DeepSeekCoder technical report \citep{guo2024deepseek} reports results on HumanEval and MBPP for the 7B-instruct-v1.5 and the 33B-instruct models, the report additionally report DS1000 results for the 33B-instruct model.
We could reproduce the original results on HumanEval and DS-1000, but got slightly worse results on MBPP because they used few-shot prompting, which should outperform our zero-shot method. 

\noindent \textbf{Llama3} \quad
Since there is no technical report available yet, the official blog post \footnote{\url{https://ai.meta.com/blog/meta-llama-3/}} report results on HumanEval, without any descriptions on prompting construction or the inference process. Our reproduced results are about 4 points lower than their original results.

\section{Analysis on Open-Domain Coding Problems}
\label{app:open-domain-study}

In \S\ref{sec:3.4:crag-results}, providing the documentation of required libraries brings limited benefits, especially with strong proprietary models such as GPT and Gemini. While we hypothesize that these strong models are sufficiently familiar with the required libraries and in turn barely benefit from additional information about them, in this section, we quantitatively investigate this issue and verify its validity.

Concretely, we construct a subset of ODEX containing only examples with less common libraries. We use the real-world distribution of all libraries involved in ODEX and select examples that use the top 20 least common libraries (e.g., \texttt{sqlite3}, \texttt{ftplib}, \texttt{flask}). We then evaluate model performance on this subset and compare the results with and without documentation in model contexts.

\noindent \textbf{With varied retrieval models} \quad
Aligning with \S\ref{sec:3.4:crag-results}, we examine the RACG results using documentation retrieved by different retrieval models. As shown in \autoref{tab:odex-study-gpt}, augmenting documentation retrieved with most methods improves the results by 20.3--40.1\%.
Compared to the entire ODEX set where most queries require common libraries, this hard-library split more clearly demonstrates the effectiveness of augmenting library documentation. This result verifies our hypothesis that strong GPT models are familiar with most common libraries, and can only benefit from additional library information when harder libraries are required.

\begin{table}[ht]
\vspace{-2mm}
\small
\centering
\resizebox{0.50\textwidth}{!}{
  \begin{tabular}{l|c|cccc|c}
    \toprule
    \textbf{Model} & {none} & {BM25} & {GIST} & {Voyage} & {OpenAI} & \gtcell{Gold} \\
    \midrule
    {GPT-3.5-turbo} & {17.2} & \good{\bf 24.1} & \bad{13.8} & \good{20.7} & {17.2} & \good{\bf 24.1} \\
    {GPT-4} & {20.7} & \good{24.1} & \bad{17.2} & \good{27.6} & \good{24.1} & \good{27.6} \\
    \bottomrule
  \end{tabular}
}
\vspace{-1mm}
\caption{RACG results on the subset of ODEX examples using the least common libraries.}
\label{tab:odex-study-gpt}
\vspace{-2mm}
\end{table}

\section{How Many Documents to Augment?}
\label{app:num-docs}
Different models have varied context length limits and context utilization abilities. Therefore, we study how model performance varies when providing different numbers of documents in the context. 
We experiment with one representative dataset for each task category: HumanEval since it is the most commonly used dataset, ODEX for its broad domain coverage, and RepoEval for its solvable difficulty.
We compare RACG performance when providing top-1, 2, 5, and 10 documents.

\begin{figure*}[ht]
\vspace{-1mm}
    \centering
    \includegraphics[width=\textwidth]{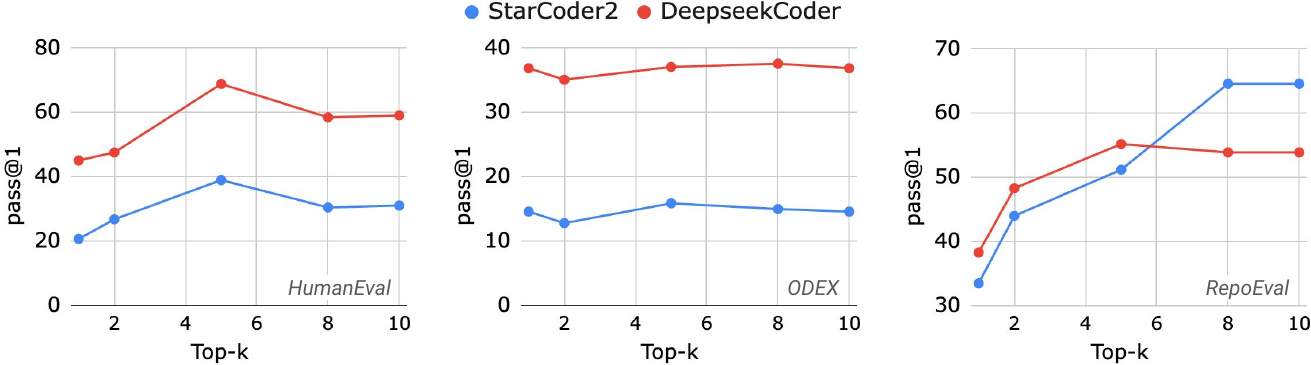}
    \vspace{-6mm}
    \caption{Comparing RACG performance with various numbers of documents.}
    \label{fig:compare-topk}
\vspace{-1mm}
\end{figure*}

As shown by \autoref{fig:compare-topk}, adding five documents yields the best results in most settings, except for StarCoder2 on RepoEval which best uses 8 documents. 
Despite the drastic variance in length limits of StarCoder2 (16$k$) and DeepseekCoder (4$k$), the sweet spot is consistently 5 documents. While adding a few documents brings helpful contexts, adding more low-ranked documents may introduce noise and deteriorate generation due to the imperfections of retrieval systems \citep{wang2023learning}.
\section{Does RACG Help Stronger Models?}
\label{app:open-racg-strong-lm}
We have shown that RACG with open retrieval improves a relatively weaker model, StarCoder2 (\S\ref{sec:4:retrieve-sources}). To see if this improvement of RACG with open retrieval generalizes to stronger models, we experiment with a series of top-performing proprietary models: GPT-4o, Claude-3-haiku and sonnet, and Gemini-1.5-flash and pro.

\noindent \textbf{Basic programming: HumanEval} \quad
RACG can consistently improve the performance of GPT-4 and Claude-3-sonnet when leveraging all sources of documents. However, for weaker models such as Claude-3-haiku and Gemini-1.5-flash, RACG only helps when aggregating multiple sources yet falls short when grounding on one source (even the canonical solution source).
Interestingly, the stronger Claude-3-sonnet performs worse than the weaker Claude-3-haiku, but can benefit from all retrieval sources and outperform haiku with documents from the canonical programming source, suggesting its potentially better RAG ability.
While the stronger Claude effectively benefits from extra contexts, the stronger Gemini-1.5-pro behaves similarly to its weaker counterpart and cannot do RACG effectively with non-canonical sources.

\vspace{1mm}
\noindent \textbf{Open domain: ODEX} \quad
All models experience limited improvements by leveraging library documentation to complex the ODEX task, with the only exception that GPT-4o improves $4.6$ points by incorporating programming solutions into the context. 

As results degrade in most cases, we conduct a manual analysis to examine when most models fail. We find that most models tend to copy functions in the context, sometimes even overwriting the function being queried, thus failing all the test cases specific to the queried function. Further, possibly affected by the plethora of programs in context, models tend to generate over-complicated programs which, however, do not often pass the test cases. 

In general, most models can be easily distracted or disturbed by additional contexts \citep{wang2023learning}, and fail to conduct the designated code generation task, indicating much room for improvement for RACG.

\begin{table*}[t!]
\vspace{-2mm}
\small
\centering
\resizebox{0.7\textwidth}{!}{
  \begin{tabular}{l|c|ccccc|c}
    \toprule
    \textbf{Method} & {Baseline} & \dlcell{Program} & {Tutorial} & {Docs} & {SO} & {GitHub} & {All} \\
    \midrule
    {GPT-4o} & {75.6} & \dlcell{94.5} & \better{90.2} & \better{90.9} & \better{91.5} & \good{84.8} & \better{95.1} \\
    \midrule
    {Claude-3-haiku} & {74.4} & \dlcell{77.4} & \good{77.4} & \bad{71.3} & \bad{67.7} & \bad{73.2} & \good{82.9} \\
    {Claude-3-sonnet} & {65.9} & \dlcell{78.7} & \good{66.5} & \good{68.9} & \good{70.7} & \good{73.8} & \better{80.5} \\
    \midrule
    {Gemini-1.5-flash} & {72.0} & \dlcell{91.5} & \good{75.0} & \bad{70.1} & \bad{68.9} & \bad{68.9} & \better{95.1} \\
    {Gemini-1.5-pro} & {82.9} & \dlcell{95.7} & \bad{79.9} & \bad{77.4} & \bad{79.9} & \bad{80.5} & \good{86.6} \\
    \bottomrule
  \end{tabular}
}
\vspace{-1mm}
\caption{RACG on HumanEval with strong code LMs.}
\label{tab:humaneval-racg-strong}
\vspace{-1mm}
\end{table*}

\begin{table*}[t!]
\vspace{-2mm}
\small
\centering
\resizebox{0.7\textwidth}{!}{
  \begin{tabular}{l|c|ccccc|c}
    \toprule
    \textbf{Method} & {Baseline} & {Program} & {Tutorial} & \dlcell{Docs} & {SO} & {GitHub} & {All} \\
    \midrule
    {GPT-4o} & {44.6} & \good{49.2} & \bad{44.2} & \dlcell{47.6} & \bad{40.3} & \bad{39.4} & \bad{39.6} \\
    \midrule
    {Claude-3-haiku} & {48.5} & \bad{42.6} & \bad{39.2} & \dlcell{44.6} & \worse{33.7} & \bad{40.5} & \bad{35.1} \\
    {Claude-3-sonnet} & {41.0} & \bad{37.6} & \bad{35.3} & \dlcell{38.0} & \bad{34.2} & \good{42.4} & \bad{38.0} \\
    \midrule
    {Gemini-1.5-flash} & {50.6} & \bad{48.3} & \bad{46.7} & \dlcell{46.2} & \bad{41.9} & \bad{44.9} & \bad{43.1} \\
    {Gemini-1.5-pro} & {57.2} & \bad{54.4} & \worse{45.6} & \dlcell{51.0} & \worse{46.5} & \worse{39.6} & \worse{46.0} \\
    \bottomrule
  \end{tabular}
}
\vspace{-1mm}
\caption{RACG on ODEX with strong code LMs.}
\label{tab:odex-racg-strong}
\end{table*}

\vspace{1mm}
\noindent \textbf{Repository level: RepoEval} \quad
While GPT-4o can solve the RepoEval task with a reasonable success rate, all Claude models are challenged by the task and achieve less than 10\% pass@1 for most scenarios. We find Claude models mostly respond with explanations of the incomplete input code, instead of the to-be-completed code even with proper instructions, possibly caused by some properties of the unknown training data. 
Gemini-1.5-flash also barely solves the task and often generates textual explanations; however its stronger pro variant gets about 10--25 point improvements, demonstrating its stronger repository-level code completion abilities.

\begin{table*}[t!]
\vspace{-2mm}
\small
\centering
\resizebox{0.8\textwidth}{!}{
  \begin{tabular}{l|cc|ccccc|cc}
    \toprule
    \textbf{Method} & {Baseline} & \dlcell{Local} & {Program} & {Tutorial} & {Docs} & {SO} & {GitHub} & {All} & {L+E} \\
    \midrule
    {GPT-4o} & {32.4} & \dlcell{62.2} & \good{35.4} & \bad{28.7} & \bad{27.8} & \bad{29.0} & \bad{28.2} & \bad{30.3} & \better{54.2} \\
    \midrule
    {Claude-3-haiku} & {~~9.1} & \dlcell{~~0.5} & \bad{~~0.5} & \bad{~~0.5} & \bad{~~0.5} & \bad{~~0.5} & \bad{~~0.2} & \bad{~~0.2} & \bad{~~0.5} \\
    {Claude-3-sonnet} & {~~0.5} & \dlcell{~~0.5} & \bad{~~0.5} & \bad{~~0.5} & \bad{~~0.5} & \bad{~~0.5} & \bad{~~0.5} & \bad{~~0.5} & \bad{~~0.5} \\
    \midrule
    {Gemini-1.5-flash} & {~~1.3} & \dlcell{16.9} & \good{~~4.0} & \good{~~2.1} & \good{~~3.2} & \good{~~2.1} & \good{~~3.2} & \good{~~2.7} & \better{11.8} \\
    {Gemini-1.5-pro} & {10.5} & \dlcell{39.1} & \good{15.1} & \good{13.4} & \good{15.8} & \good{15.3} & \good{11.8} & \good{12.3} & \better{33.0} \\
    \bottomrule
  \end{tabular}
}
\vspace{-1mm}
\caption{RACG on RepoEval with strong code LMs.}
\label{tab:repoeval-racg-strong}
\vspace{-2mm}
\end{table*}

\end{document}